\documentclass[a4paper,10pt]{article}
\usepackage{graphicx}
\usepackage{natbib,amssymb}

\usepackage{amsthm}
\usepackage{amsmath}
\usepackage[latin1]{inputenc}
\usepackage{epsfig}
\usepackage{epstopdf}
\usepackage{fancybox}
\usepackage{float}
\usepackage{pifont}
\usepackage{fancyhdr}
\usepackage{ulem}
\usepackage{url}
\usepackage{color}
\usepackage{placeins}  %Pour mettre les barrierres a flottants
\usepackage{setspace}  % Pour double spacer
\usepackage{lastpage}

\def \RR {\mathbb{R}}

\def \N  {\mathcal{N}}

\def\bx{\boldsymbol{x}}
\def\by{\boldsymbol{y}}
\def\bz{\boldsymbol{z}}

\def\bh{\boldsymbol{h}}
\def\bw{\boldsymbol{w}}
\def\br{\boldsymbol{r}}
\def\bepsilon{\boldsymbol{\epsilon}}
\def\btheta{\boldsymbol{\theta}}
\def\bbeta{\boldsymbol{\beta}}
\def\I{{\bf I}}
\def\H{{\bf H}}
\def\W{{\bf W}}
\def\R{{\bf R}}

\def\Q {{\bf Q}}
\def\U{{\bf U}}
\def\S{{\bf S}}
\def\V{{\bf V}}
\def\bPi{{\bf \Pi}}

\DeclareMathOperator*{\argmin}{arg\,min}

\title{ Calibration and improved prediction of computer models by universal Kriging }
\author{ Fran\c cois Bachoc\thanks{Corresponding author.
E-mail: francois.bachoc@cea.fr Address: CEA-Saclay, DEN, DM2S, STMF, LGLS, F-91191 Gif-Sur-Yvette, France.
Phone: +33 (0) 1 69 08 97 91} \\
{ \small CEA-Saclay, DEN, DM2S, STMF, LGLS, F-91191 Gif-Sur-Yvette, France} \\
{ \small Laboratoire de Probabilit\'es et Mod\`eles Al\'eatoires, Universit\'e Paris VII} \\
{ \small Site Chevaleret, case 7012, 75205 Paris cedex 13}  \\
\\
Guillaume Bois \\
{ \small CEA-Saclay, DEN, DM2S, STMF, LATF, F-91191 Gif-Sur-Yvette, France} \\
\\
Josselin Garnier \\
{ \small     Laboratoire de Probabilit\'es et Mod\`eles Al\'eatoires \& Laboratoire Jacques-Louis Lions} \\
 { \small Universit\'e Paris VII, Site Chevaleret, case 7012, 75205 Paris cedex 13, France } \\
\\
Jean-Marc Martinez \\
{ \small CEA-Saclay, DEN, DM2S, STMF, LGLS, F-91191 Gif-Sur-Yvette, France} }

\date{}

\begin{document}

\maketitle

{ \large \textbf{Total number of pages: \pageref{LastPage}}}

\listoffigures  % table des figures
\listoftables   % table des tableaux

\newpage

\doublespacing

\begin{abstract}
This paper addresses the use of experimental data for calibrating a computer model
and improving its predictions of the underlying physical system.
A global statistical approach is proposed in which
the bias between the computer model and the physical system is modeled as a realization of a Gaussian process. 
The application of classical statistical
inference to this statistical model yields a rigorous method for
calibrating the computer model and for adding to its predictions a statistical correction based on experimental data.
This statistical correction can substantially improve the calibrated computer model for predicting the physical system on new experimental conditions.
Furthermore, a quantification of the uncertainty of this prediction is provided.
Physical expertise on the calibration parameters can also be taken into account in a Bayesian framework.
Finally, the method is applied to the thermal-hydraulic code FLICA 4, in a single phase friction model framework.
It allows to improve the predictions of the thermal-hydraulic code FLICA 4 significantly.
\end{abstract}

\newpage

\section{Introduction} \label{section: introduction}

In physics and engineering, computer experiments have been used for a long time as surrogates to costly, or unpractical, real experiments. However, the
impact of the substitution of a computer experiment to an actual experiment is not easy to assess. In general, this substitution induces a double error.
The first error is that
a computer experiment is the numerical implementation of a mathematical model, so that a bias always exists between this mathematical model and the computer model.
The quantification and reduction of this first error is the field of model verification \cite{VCCCSE}. The second error is that the mathematical model itself may not
represent perfectly the underlying physical phenomenon. This second error defines the field of model validation. A reference book on model validation is e.g. \cite{SUAT}.

In the present work, and similarly to several references on statistical analysis of the validation problem \cite{BVCM},
we assume that the computer model has already been verified.
Hence, we focus on the validation problem which is a very important issue in nuclear engineering \cite{THSCNRSQP}.
Nevertheless, our objective is less to study the validity of the computer model than to improve the computer model predictions,
and quantify the uncertainty obtained, by assimilating experimental results.
A recent reference on demonstrating, or refuting, the validity of the actual computer model would rather be \cite{BVCM}.

In practice, the study of computer model predictions is complicated by the fact that a computer model often comes with fitting parameters.
These parameters either allow to model a physical system more accurately or
are a consequence of uncertainties with respect to some physical parameters. 
Hence, the analysis of the predictions of the computer model often needs to be carried out simultaneously with a calibration
analysis. The goal of the calibration analysis is to quantify the uncertainty related to the fitting parameters.
An example of a methodology carrying out
calibration and prediction analysis globally is the Best-Estimate methodology \cite{BEMCPTEDAMF,BEMCPTEDAABBE}.
This methodology proceeds through data assimilation and takes into account the different
sources of uncertainties coming from the model parameters, the experimental errors, the numerical errors, and the limitations of the physical models.

In this work, a computer model is a function $f_{mod}$ of the form
$f_{mod}(\bx,\bbeta) :  \RR^d \times \RR^m  \rightarrow \RR$.
This computer model is a representation of a physical system that is a deterministic function taking the form
$f_{real}(\bx) :   \RR^d  \rightarrow \RR$.

The scalar output of the physical system is the physical quantity of interest.
It is a deterministic function $f_{real}$ of a vector $\bx$ of
input quantities, that we call \textbf{experimental conditions}.
These components of the vector $\bx$ can be divided into two categories.
The first category contains the \textbf{control variables}. These variables define the physical system, independently of the environment in which
the system is put. In engineering for instance, geometric parameters of the system can often be placed in this category, since they remain fixed regardless of
what happens to the system.
The second category contains the \textbf{environment variables}. These variables are the input of the physical system which values are not planned in the conception
of the system. These variables are likely to be imposed beforehand by other systems. 
The distinction of the experimental conditions into these two categories is presented for instance in \cite{TDACE} section 2.1. To give an illustration, in the system design
phase, the environment variables are set by the future use of the system, while the control variables are the free parameters that may be set through an optimization phase.

The function $f_{real}$ of the physical system can not be evaluated for all the experimental conditions.
Hence, this function is approximated by the computer model $f_{mod}$.
This function shares the same input vector $\bx$ as the physical system and provides the same scalar output. Furthermore, the function $f_{mod}$ has a second kind of inputs,
denoted by the vector $\bbeta$. The components of this vector are the fitting parameters of the computer model $f_{mod}$.
These parameters are unnecessary to carry out an experiment
of the physical system, but they are needed to run the computer model. Hence, these quantities are seen as
degrees of freedom for the computer model, and allow it to give a good approximation of the physical system. In the sequel these parameters are called \textbf{model parameters}.

The calibration problem is the problem of computing a value $\hat{\bbeta}$ for the model parameter $\bbeta$. This is done on the basis of a set of experimental
results. A set of experimental results is a set $\left\{ \bx^{(1)},f_{obs}(\bx^{(1)}),...,\bx^{(n)},f_{obs}(\bx^{(n)}) \right\}$ where
$f_{obs}(\bx^{(i)})$ is the output of the physical system observed at the experimental condition $\bx^{(i)}$. Note that the quantity
$f_{obs}(\bx^{(i)})$ may be different from $f_{real}(\bx^{(i)})$, for example because of measurement errors.
In practice, in nuclear engineering, calibration is an important issue because many computer models usually have model parameters that are totally or
partially unknown \cite{BCCM}.

Before presenting the Gaussian process modelling of this work, we emphasize the potential limitations of
uncertainty quantification methods that would only address calibration. The most classical example of
these methods is the least square calibration, which consists in
minimizing with respect to $\bbeta$ a quadratic misfit between the experimental results and the values that the computer model predicts when
parameterized by $\bbeta$. Hence, the value of $\bbeta$ obtained with the least square method is
$\hat{\bbeta}_{LS} \in \argmin_{\bbeta} \sum_{i=1}^n  \left( f_{mod}(\bx^{(i)},\bbeta) - f_{obs}(\bx^{(i)}) \right)^2$.
A method addressing only the calibration problem, such as the least square calibration method, is generally based on two hypotheses.
The first hypothesis is that the computer model is capable to perfectly reproduce the physical system. That is to say, there is a model parameter $\bbeta_0$ so that
$\forall \bx, f_{real}(\bx) = f_{mod}(\bx,\bbeta_0)$. This hypothesis means that the physical knowledge that was put in the computer model is sufficient to
model the physical system perfectly. 
The second hypothesis is that the deviations $(f_{mod}(\bx^{(i)},\bbeta_0) - f_{obs}(\bx^{(i)}))$ come from uncertainties related to the experiments.
These uncertainties have generally two sources.
First, the observations are affected by measurement errors. Second, there is a replicate uncertainty, meaning that the experimental conditions can not be known exactly
for a given experiment. 
The main limitation is the assumption that the deviations
$f_{mod}(\bx^{(i)}, \bbeta) - f_{obs}(\bx^{(i)})$ come only from uncertainties related to the experiments.
Indeed the order of magnitude of these uncertainties is known.
Hence, when the errors $|f_{mod}(\bx^{(i)}, \bbeta) - f_{obs}(\bx^{(i)})|$ are too large compared to this order of magnitude, it indicates that there
is a problem with this assumption (this can be quantified by Monte Carlo methods).
In this case, the fact that the computer model can not represent the physical system perfectly
needs to be taken into account. 

The limitation discussed above give motivations for reconsidering the
assumption that there is a model parameter $\bbeta_0$ so that
$\forall \bx, f_{real}(\bx) = f_{mod}(\bx,\bbeta_0)$.
In this paper, this is done by taking  a model error into account.

In Section \ref{section: GaussianProcessValidationMethod}, we present in detail the Gaussian process modelling of the model error,
and show how this modelling yields a framework for calibration and prediction. We also give a one-dimensional illustration
on an analytical function. In Section \ref{section: application}, we present an application case, relevant to Nuclear Engineering, on the thermal-hydraulic code FLICA 4.
The thermal-hydraulic code FLICA 4 is mainly dedicated to core thermal-hydraulic transient and steady state analysis \cite{FDPFCCANMNA}.

\section{ The Gaussian process method for calibration and prediction} \label{section: GaussianProcessValidationMethod}

\subsection{The Gaussian process modelling of the model error}

We present the Gaussian process model, that is the basis of the Gaussian process method for calibration and prediction.
This statistical model is based on two main ideas:
\begin{itemize}
\item The physical system $ \bx \to f_{real}(\bx)$ does not necessarily belong to the set of computer model functions
$\{ f_{mod}( \bx , \bbeta) \}$. We model the difference between the physical system and the correctly parameterized computer model by an error function
that is called the \textbf{model error}. The notion of correctly parameterized computer model is explained below. 
\item The model error function is not observable everywhere, and hence is unknown for the majority of the experimental conditions. This lack of
knowledge is modeled by the introduction of a stochastic framework for this function, that is to say it is represented by a realization of the random process $Z(\omega,\bx)$.
This probabilistic
modelling is a Bayesian modelling of the uncertainty on the deterministic model error function. The reader may refer to \cite{TDACE} $p 23,24$ for a discussion
of Bayesian modelling of deterministic functions.
In this context, the particular interest of Gaussian processes is discussed in \cite{GPML} $p2$. Being the sum of the correctly
parameterized computer model and of the model error function, the physical
system itself is a realization of a Gaussian process. 
Hence, we do not use the notation $f_{real}(\bx)$ anymore for the physical system.
Instead, we denote it by the random process $Y_{real}(\omega,\bx)$.   

\end{itemize}

Motivated by these two ideas, the Gaussian process statistical model is defined by the two following equations

\begin{equation} \label{eq: mod_proc}
Y_{real}(\omega,\bx) = f_{mod}(\bx,\bbeta) + Z(\omega,\bx) 
\end{equation}
and
\begin{equation} \label{eq: mod_obs}
Y_{obs}(\omega,\bx) = Y_{real}(\omega,\bx) + \epsilon(\omega,\bx). 
\end{equation}
With:
\begin{itemize}
\item $\omega$ in a probability space $\Omega$.
\item $Y_{real}(\omega,\bx)$ the random process of the physical system.
\item $Z(\omega,\bx)$ is the model error process. The random process $Z$ is assumed to be Gaussian \cite{TDACE,ISDSTK,GEUKUP} and centered.
The model error process $Z$ is hence defined by its covariance
function $C_{mod}$. In practice, this function belongs to a prescribed set of covariance functions (see e.g \cite{RGRFCF} for classical examples) and is defined up to
a few hyper-parameters that can be estimated from data. 
\item $\bbeta$ is the correct parameter of the computer model. We call it the correct parameter because, $Z$ being centered, the computer model parameterized by $\bbeta$ is
the mean value of the physical system.
\item $Y_{obs}(\omega,\bx)$ is the observed output of the physical system for the experimental conditions $\bx$. This observation is the sum of
the quantity of interest and of a measurement error $\epsilon(\omega,\bx)$. $\epsilon(\omega,\bx)$ follows a Gaussian centered law, and is independent from
one experiment to another. The variance of $\epsilon$ is in general constant.
\end{itemize}

In the sequel we do not write explicitly the $\omega$.
One could model $Z$ as a white noise process (with independent components for different $\bx$), which would give a statistical model leading to
the least square calibration of section \ref{section: introduction}. There are two reasons for
not doing so, and use instead a covariance function with a dependence structure.
\begin{itemize}
\item The physical system is generally continuous with respect to the experimental conditions, and so is the numerical model. Hence, as a difference,
the model error process $Z$ must be a process with continuous trajectories. This is not the case for a white noise process.
\item Similarly, it is expected that if the computer model makes a certain error for a given experimental point, then it will do a similar error for a nearby experimental
point. This principle is taken into account by a covariance function with a dependence structure.
\end{itemize}

The statistical modelling also allows to take into account expert judgments for the model parameter $\bbeta$. This is done within the
Bayesian framework, modelling the constant but unknown correct model parameter $\bbeta$ as a random vector. The law of this random vector is known, and
chosen according to the degree of knowledge one has about the model parameter $\bbeta$. We use a Gaussian distribution for the Bayesian modelling of
$\bbeta$.
Hence, we distinguish two cases:

\begin{description}
\item{ \textbf{No prior information case:} } $\bbeta$ is a vector of unknown constants.  
\item{ \textbf{Prior information case:} } $\bbeta$ is a random vector, with known mean vector $\bbeta_{prior}$ and covariance matrix $\Q_{prior}$. 
\end{description}

In this paper, we work with a linear approximation of the computer model with respect to its model parameters (within the range of values that is under consideration).
Hence we consider computer models of the form
$ f_{mod}(\bx,\bbeta) = f_{mod}(\bx,\bbeta_{nom}) + \sum_{i=1}^m h_i(\bx) (\bbeta_i - \bbeta_{nom,i} )$
where $\bbeta_{nom}$ is the nominal vector around which the linear approximation is made.
$\bbeta_{nom}$ is generally chosen by expert judgment or by previous calibration studies.
We
choose, for simplicity reasons, to remove
the perfectly known quantities $\bbeta_{nom}$ and $f_{mod}(\bx,\bbeta_{nom})$. Indeed, up to a shift with respect to $\bbeta$ and $f_{mod}$, we can consider
that $\bbeta_{nom} = 0$ and $f_{mod}(\bx,\bbeta_{nom}) = 0$. We then have
\begin{equation} \label{eq: linear_approximation}
\forall \bx:~ f_{mod}(\bx,\bbeta) = \sum_{i=1}^m h_i(\bx) \bbeta_{i}.
\end{equation}

The linear approximation is justified by a Taylor series expansion when the uncertainty concerning the correct parameter $\bbeta$ is small.
This linear approximation is frequently made, for example in thermal-hydraulics \cite{deCrecy01,BEMCPTEDAABBE}, or in neutron transport \cite{kawano2006}. 
A thorough discussion on the validity of using the linear approximation in the non-linear case is given in section \ref{subsection: recommendation}.

The Gaussian process modelling allows to solve the two following problems:
\begin{enumerate}
\item \textbf{Calibration.} It is the problem of estimating the correct model parameter $\bbeta$, or equivalently to find the most accurate computer model function
$\bx \to f_{mod}(\bx,\bbeta)$. 
\item \textbf{Prediction.} For a new experimental condition $\bx_{new}$, we want to predict the quantity of interest of the physical system, and add a measure of
uncertainty to this prediction. The main idea is that the quantity of interest is not predicted by the calibrated computer model, because we are able to
infer the value of the model error at $\bx_{new}$.
\end{enumerate}
The calibration and prediction are presented in section \ref{subsection: calPred}. They are obtained using classical linear algebra tools, as long as the
covariance function $C_{mod}$ of the model error is known.
In fact, the function $C_{mod}$ depends on a set of hyper-parameters that are
to be estimated from data. We present the estimation method in \ref{subsection: RMLE}. 

Hence, in the most classical case, the Gaussian process modelling in treated in two steps. In a first step, the hyper-parameters of the covariance function are
estimated so that this function is considered fixed in the second step, where linear algebra is used for the calibration and prediction.
Note that there exist methods where these two steps
are done simultaneously, for instance if a Bayesian prior for the hyper-parameters of the covariance function is used. These methods are more costly, but can
improve the quality of the Gaussian process modelling, as shown e.g in \cite{BOUSMC} in an optimization context.  

Note that in our case, a third step of verification is necessary. This step consists in verifying that the modelling leads to calibration and prediction that
give satisfying results. This step can notably be carried out by \textbf{Cross Validation}, or using a new set of experimental results that was never used before.

We now formulate the problem in vector-matrix form.
Assume that $n$ experiments are carried out at $\bx^{(1)},...,\bx^{(n)}$.
We denote:
\begin{itemize}
\item The $n \times m$ matrix $\H$ of partial derivatives of the computer model with respect to $\bbeta = (\beta_1,...,\beta_m)$. $\H$ is defined by $H_{i,j} = h_j(\bx^{(i)})$.
\item The random vectors $\by_{obs}$ of the $n$ observations. $y_{obs,i}$ is the result of the $i$th experiment.
\item The random vector $\bepsilon$ of the measurement errors for the $n$ experiments. We have $y_{obs,i} = Y_{real}(\bx^{(i)}) + \epsilon_i$.
\item The random vector $\bz$ of the model error for the $n$ experiments. $\bz = (Z(\bx^{(1)}),...,Z(\bx^{(n)}))^t$.
\item The covariance matrix of $\bepsilon$: $\R_{mes}$. 
\item The covariance matrix of $\bz$: $\R_{mod}$. 
\end{itemize}
For the $n$ experiments the equations \eqref{eq: mod_proc} and \eqref{eq: mod_obs} become
\begin{equation} \label{eq: modelFiniteSample}
\by_{obs} = \H \bbeta + \bz + \bepsilon.
\end{equation}
Hence we have a universal Kriging model \cite{TDACE}. We denote by $\R$ the covariance matrix of the model and measurement error vector $\bz + \bepsilon$.
\begin{equation}  \label{eq: RmodK}
\R := cov(\bz + \bepsilon) = \R_{mod} + \R_{mes}, 
\end{equation}
The distinction between $\R_{mod}$ and $\R_{mes}$ is important because, usually $\R_{mes}$ is at least partially known from knowledge of the experimental process while $\R_{mod}$
generally does not benefit from physical knowledge. Indeed, physical knowledge is used in the conception of the computer model, and hence may not help knowing the shape of
the error of the computer model.

We can compute the {\it a priori} law of the vector of observations, i.e. the statistical distribution of the observations before carrying the experiments,
but given the hyper-parameters.
In the no prior information case we have, with $\bbeta$ an unknown constant,
\begin{equation} \label{eq: vect_gauss_siap}
 \by_{obs} \sim \N( \H \bbeta , \R ). 
\end{equation}
In the prior information case, we have, with $\bbeta \sim \N(\bbeta_{prior},\Q_{prior})$
\begin{equation} \label{eq: vect_gauss_aiap}
 \by_{obs} \sim \N( \H \bbeta_{prior} , \H \Q_{prior} \H^t + \R ). 
\end{equation}
Here $\N(\boldsymbol{m},\R)$ stands for the multivariate Gaussian distribution with mean vector
$\boldsymbol{m}$ and covariance matrix $\R$ and $\sim$ means "follows the distribution of".

\subsection{Estimation of the covariance of the model error} \label{subsection: RMLE}

For the Gaussian process modelling defined in \eqref{eq: mod_proc} and \eqref{eq: mod_obs} to be tractable with closed form linear algebra formulas
(subsection \ref{subsection: calPred}), it is necessary that the covariance functions of the model error $Z$ and of the measurement error $\epsilon$ are known.
We show here how to compute these covariance functions.

The covariance function of the measurement error can generally be specified form physical expertise. This is the case here. If it is not the case, this function
can, for example, be estimated in the same way as the model error covariance function.

Generally there is no expert judgment available concerning the model error covariance function $C_{mod}$, as has been discussed above. A specific structure is chosen
for $C_{mod}$, with a limited number of degrees of freedom. Hence we consider the family of covariance functions
\[
\mathcal{C}_{mod} = \left\{ \sigma^2 C_{mod,\btheta} , \sigma >0 , \btheta \in \Theta \right\}.   
\]
Here $\Theta$ is a subset of $\RR^p$ and $C_{mod,\btheta}$ is a stationary correlation function. We present classical correlation functions families, for
which $p=d$, $\btheta = ( l_{c,1},...,l_{c,d})$ and $C_{mod,\btheta} (\bx^{(a)},\bx^{(b)}) = C_{mod,\btheta} (\bh)$ with $\bh = \bx^{(a)} - \bx^{(b)}$. The component
$l_{c,i}$ can be seen as a correlation length in the $i$th dimension.

\begin{itemize} \label{itemize: covarianceFunctions}
\item exponential correlation function
\[
C_{mod,\btheta}(\bh) = \exp{ ( - \sum_{i=1}^d \frac{|h_i|}{l_{c,i}} ) }.
\]
\item Mat\'ern $\nu = \frac{3}{2}$ correlation function, with $|\bh|_{\btheta} = \sqrt{ \sum_{i=1}^d \frac{h_i^2}{l_{c,i}^2} }$
\[
C_{mod,\btheta}(\bh) = ( 1 + \sqrt{6} |\bh|_{\btheta} )  \exp{ ( - \sqrt{6} |\bh|_{\btheta} ) } 
\]
\item Mat\'ern $\nu = \frac{5}{2}$ correlation function
\[
C_{mod,\btheta}(\bh) = ( 1 + \sqrt{10} |\bh|_{\btheta} + \frac{10}{3} |\bh|_{\btheta}^2 )  \exp{ ( - \sqrt{10} |\bh|_{\btheta} ) } 
\]
\item Gaussian correlation function
\[
 C_{mod,\btheta}(\bh) = \exp{ ( - |\bh|_{\btheta}^2 ) }
\]
\end{itemize}
The correlation functions above yield sample functions of increasing regularity (see for instance \cite{ISDSTK}).
The importance of the regularity of the correlation function is presented in detail in \cite{ISDSTK}.  

Assume now that we have $n$ experimental results $\by_{obs} = (y_{obs,1},...,y_{obs,n})$ and recall the notations of \eqref{eq: modelFiniteSample} and \eqref{eq: RmodK}. As we have seen, $\R_{mes}$ is fixed, and
$\R_{mod}$ depends on $(\sigma^2,\btheta)$ that are to be estimated. We use the notation $\R_{\sigma,\btheta}$ for the global covariance matrix $\R = \R_{mod} + \R_{mes}$.

There are several methods that can be used to estimate the hyper-parameters $(\sigma^2,\btheta)$
from data $\by_{obs}$. The most widely used are Maximum Likelihood \cite{MLEMRCSR} and Cross
Validation \cite{KCVMSD,PACHGP}. 

In this work, we use the Restricted Maximum Likelihood Estimator (RMLE) of $(\sigma^2,\btheta)$. This estimator is for instance presented in \cite{ADREMLE}.
The advantage of this estimator is that the estimation of $(\sigma^2,\btheta)$ is independent of the estimation of $\bbeta$. Furthermore, this method allows to
have the same estimation of $(\sigma^2,\btheta)$ in both the prior and no prior information case. Finally, let us notice that $n>m$ is required for the REML method,
that is to say there are more experiments that model parameters. In thermal-hydraulics, the field of the application case, this condition generally holds in practice.
Nevertheless, in other fields of Nuclear Engineering, typically in neutron transport \cite{kawano2006}, one may have $m>>n$. In this case, if one want to
address the present model error modelling anyway, it is recommended to work in a fully Bayesian framework, both for the model parameters and
the covariance hyper-parameters
as described in \cite{TDACE} section 4.1.4. Indeed, the very large number of model parameters makes the uncertainty related to the
hyper-parameters of the model error covariance function too large to be neglected, as is done when these hyper-parameters are fixed to their estimated values.

Let $\W$ be a $(n-m \times n)$ matrix of full rank so that $\W \H = 0$. Notice that if $\H$ is not of full rank,
then $m$ must be replaced by the rank of $\H$.
Then
\[
\bw := \W \by_{obs} \sim \N( 0 , \W \R_{\sigma , \btheta} \W^t ). 
\]
The law of $\bw$ is independent of the value of $\bbeta$. Hence the RMLE $(\hat{\sigma}^2,\hat{\btheta})$ is the Maximum Likelihood estimator on the transformed
observations $\bw$:
\begin{equation} \label{eq: REML}
(\hat{\sigma} , \hat{\btheta} ) \in \argmin_{ (\sigma , \btheta) } q(\sigma,\btheta)
\end{equation}
with:
\begin{equation} \label{eq: REMLq}
q(\sigma,\btheta) =  \ln{ | \W \R_{\sigma,\btheta} \W^t | } + \bw^t ( \W \R_{\sigma,\btheta} \W^t  )^{-1}  \bw.
\end{equation}
It is shown in \cite{BIVCUOEC} that changing $\W$ only adds a constant (with respect to $(\sigma^2,\btheta)$) term to \eqref{eq: REMLq}.
It is also shown in \cite{BIVCUOEC} how one can avoid a matrix product with $\W$. Indeed for $\W$ so that
$\W \W^t = \I_{n-m}$ and $\W^t \W = \I_n - \H (\H^t \H)^{-1} \H^t$ we have
\begin{equation} \label{eq: Harville}
q(\sigma,\btheta) = - \ln{ | \H^t \H | } +  \ln{ | \R_{\sigma,\btheta} | } +  \ln{ | \H^t \R_{\sigma,\btheta}^{-1} \H | } +
\by_{obs}^t \bPi_{\sigma,\btheta} \by_{obs}, 
\end{equation}
with
\[
\bPi_{\sigma,\btheta} = \R_{\sigma,\btheta}^{-1} - \R_{\sigma,\btheta}^{-1} \H ( \H^t \R_{\sigma,\btheta}^{-1} \H )^{-1} \H^t \R_{\sigma,\btheta}^{-1}.
\]
Let $\U,\S,\V$ be a Singular Value Decomposition of $\H$, with $\U$ of size $n \times m$ so that $\U^t \U = \I_{m,m}$, $\S$ a
diagonal matrix of size $m$, with nonnegative numbers on the diagonal, and
$\V$ an orthogonal matrix of size $m$, so that $\H = \U \S \V^t$. Then, we can show that
\begin{eqnarray}
 q(\sigma,\btheta) & = & \ln{  \left| \U^t \R_{\sigma,\btheta}^{-1} \U \right|  } + \ln{ \left| \R_{\sigma,\btheta} \right| } + \by_{obs}^t \R_{\sigma,\btheta}^{-1} \by_{obs} \nonumber \\
  &  & -  \by_{obs}^t \R_{\sigma,\btheta}^{-1} \U ( \U^t \R_{\sigma,\btheta}^{-1} \U )^{-1} \U^t \R_{\sigma,\btheta}^{-1} \by_{obs}. \label{eq: REMLavecU}
\end{eqnarray}
 
Hence, it does not matter if $\H$ is ill-conditioned, or even singular, since its singular values are actually not used in the computation of the Restricted Likelihood.
Using \eqref{eq: REMLavecU} allows both to avoid $n\times n$ matrix multiplications and to avoid numerical issues with respect to the condition number of $\H$.

\subsection{Calibration and prediction} \label{subsection: calPred}

Throughout this subsection, we assume that the covariance function $C_{mod}$ of $Z$ is estimated and fixed and we use the classical Kriging formulas
to solve the calibration and prediction problems. The Kriging formulas, in both the Bayesian and frequentist cases, can be found in \cite{TDACE}. 
We will see that these formulas require $\R$ to be non zero, i.e there are model or measurement errors.
For consistency, we first address the case $\R=0$, which is actually straightforward.
If there is a unique $\bbeta$ so that $f_{mod}(\bx,\bbeta)$ reproduces all the experiments, then
$\bbeta$ is the correct parameter with zero associated uncertainty. If there are more than one $\bbeta$
so that $f_{mod}(\bx,\bbeta)$ reproduces all the experiments, then the computer model is redundantly parameterized
or the number of experiments is insufficient. If there is no $\bbeta$
so that $f_{mod}(\bx,\bbeta)$ reproduces all the experiments, then the assumption of no model error and no measurement error is invalidated.
In the sequel, we consider $\R$ non zero, and $\R$ invertible, which is the case for the classical covariance functions of section \ref{itemize: covarianceFunctions}.

In the no prior information case, the calibration problem is the frequentist problem of estimating the unknown parameter $\bbeta$. The maximum likelihood estimation
of $\bbeta$ is 
\begin{equation} \label{eq: calSIA}
\hat{\bbeta} = (\H^t \R^{-1} \H)^{-1} \H^t \R^{-1} \by_{obs}.
\end{equation}
This estimator is unbiased and has covariance matrix
\begin{equation} \label{eq: calSIA2}
cov( \hat{\bbeta} ) =  (\H^t \R^{-1} \H)^{-1}.
\end {equation}
We see that if there is a $\bbeta$ so that $\H \bbeta = \by_{obs}$, then we have $\hat{\bbeta}=\bbeta$. This means that,
if we are in the favorable case when the computer model can perfectly reproduce the experiments, then the Gaussian process calibration of the computer model
will achieve this
perfect reproduction, as should be expected. Finally, as the random vector $\hat{\bbeta}$ has Gaussian distribution, its covariance matrix is sufficient to yield
confidence ellipsoids for $\bbeta$.

In the prior information case, the posterior distribution of $\bbeta$ given the observations $\by_{obs}$ is Gaussian with mean vector
\begin{equation} \label{eq: calAIA}
 \bbeta_{post}  = \bbeta_{prior} + (\Q_{prior}^{-1} + \H^t \R^{-1} \H)^{-1} \H^t \R^{-1} (\by_{obs} - \H \bbeta_{prior}),
\end{equation}
and covariance matrix
\begin{equation} \label{eq: calAIA2}
 \Q_{post} =  (\Q_{prior}^{-1} + \H^t \R^{-1} \H)^{-1}.
\end{equation}
We can notice that, when $\Q_{prior}^{-1} \to 0$, then the prior information case calibration tends to the no prior information case calibration.
This is an intuitive fact, because $\Q_{prior}^{-1}$ small corresponds to a small a priori knowledge of $\bbeta$ and hence should, in the limit case, correspond to
an absence of knowledge. 

\paragraph{Remark:} The prior information case calibration of \eqref{eq: calAIA} is classically used in neutron transport \cite{kawano2006},
when the linear approximation \eqref{eq: linear_approximation} of the computer model is also made. In the reference hereabove, no model error is assumed, so that the physical system
is predicted by the calibrated computer model only. In thermal-hydraulics, which is the field of the case of application,
this hypothesis is not justified. Indeed, computer models can rely on aggregation
of correlation models that have no physical justification. We will see in the prediction formulas of \eqref{eq: predSIA}
and \eqref{eq: predAIA}, and in the application case of section \ref{section: application}, that modelling the model error
allows to significantly improve the predictions of a computer model that is only partially representative of the physical system.

We now present the prediction formulas. In the same way as the computer model, the goal of the prediction is to give the most probable value of the physical system, for a new
experimental condition, without doing a real experiment. However, this most probable value is not necessarily given by the output of the calibrated computer model, because
the model error is inferred too. We now give the notations that we use for a new experimental condition $\bx_{new}$.

\begin{itemize}
\item The random value of the physical system at $\bx_{new}$: $Y_{real}(\bx_{new})$.
\item The vector of derivatives of the computer model with respect to $\beta_1,...,\beta_m$ at $\bx_{new}$: $\bh(\bx_{new})$. Hence we have
$\left( \bh(\bx_{new}) \right)_i = h_i(\bx_{new})$.
\item The model error at $\bx_{new}$: $z_{new}$. 
\item The covariance vector of the model error between $(\bx^{(1)},...,\bx^{(n)})$ and $\bx_{new}$: $\br_{mod}(\bx_{new})$ given by
$\left( \br_{mod}(\bx_{new}) \right)_i := cov(  z_i , z_{new}  )$.
\end{itemize}

In the no prior information case, the Best Linear Unbiased Predictor (BLUP) of $Y_{real}(\bx_{new})$ with respect to the vector of observations $\by_{obs}$ is
 
\begin{equation} \label{eq: predSIA}
  \hat{y}(\bx_{new}) = 
\underbrace{ (\bh( \bx_{new} ))^t \hat{\bbeta} }_{ \mbox{Calibrated computer model} } +
\underbrace{ (\br_{mod}(\bx_{new}))^t \R^{-1} (\by_{obs} - \H \hat{\bbeta} ) }_{ \mbox{Inferred model error} }.
\end{equation}

We refer to \cite{DACE} for a detailed definition of the BLUP and the computation of the predictor in \eqref{eq: predSIA}. This predictor is composed of
the calibrated computer model and of the inferred model error. By inspection of \eqref{eq: predSIA}, the inferred model error has the following properties:
\begin{itemize}
\item $\bx_{new}$ being fixed, this term is large when the errors $\by_{obs} - \H \hat{\bbeta}$ between
the experimental results and the calibrated computer model are large. 
\item  The observations being fixed, this term is a linear combination of the components of $\br_{mod}(\bx_{new})$. These elements are usually a decreasing function
of the distance between $\bx_{new}$ and the experimental conditions $\bx^{(i)}$. Hence, if $\bx_{new}$ is far from an experimental condition $\bx^{(i)}$, then the weight of
this experimental result is small in the combination. Hence, the prediction of $Y_{real}(\bx_{new})$ is almost only composed of the calibrated computer model
when $\bx_{new}$ is far from any available experimental condition, while the model error inference term is
significant when $\bx_{new}$ is in the neighborhood of an available experimental condition (the neighborhood is defined in terms of the correlation lengths).
\end{itemize}

The mean square error of the BLUP is
\begin{eqnarray} \label{eq: predSIA2}
& \hat{\sigma}^2 ( \bx_{new} ) = C_{mod}(\bx_{new},\bx_{new}) - \br_{mod}(\bx_{new})^t \R^{-1} \br_{mod}(\bx_{new}) \\ 
& + ( \bh( \bx_{new} ) - \H^t \R^{-1} \br_{mod}(\bx_{new}) )^t ( \H^t \R^{-1} \H )^{-1} ( \bh( \bx_{new} ) - \H^t \R^{-1} \br_{mod}(\bx_{new}) ). \nonumber
\end{eqnarray}
Since only linear combinations have been used, the BLUP has Gaussian distribution and the mean square error allows to build confidence intervals.

In the prior information case, the posterior distribution of $Y_{real}(\bx_{new})$ given the observations $\by_{obs}$ is Gaussian with mean
\begin{equation} \label{eq: predAIA}
\hat{y}(\bx_{new}) = 
\underbrace{ (\bh( \bx_{new} ))^t \bbeta_{post} }_{ \mbox{calibrated computer model} } +
\underbrace{ (\br_{mod}(\bx_{new}))^t \R^{-1} (\by_{obs} - \H \bbeta_{post} ) }_{ \mbox{inferred model error} },
\end{equation}
and variance
\begin{eqnarray} \label{eq: predAIA2}
& & \hat{\sigma}^2 ( \bx_{new} ) = C_{mod}(\bx_{new},\bx_{new}) - \br_{mod}(\bx_{new})^t \R^{-1} \br_{mod}(\bx_{new}) \\
& + & ( \bh( \bx_{new} ) - \H^t \R^{-1} \br_{mod}(\bx_{new}) )^t ( \H^t \R^{-1} \H + \Q_{prior}^{-1} )^{-1} ( \bh( \bx_{new} ) - \H^t \R^{-1} \br_{mod}(\bx_{new}) ). \nonumber
\end{eqnarray}
We can make the same remarks as for \eqref{eq: predSIA}. Similarly to calibration, the limit when $\Q_{prior}^{-1} \to 0$ of the prediction
in the prior information case is the prediction in the no prior information case.

\subsection{ Illustration on an analytical test case }

We illustrate the calibration and the prediction on an analytical test case. This test case is academic and allows to understand the most important features
of the Gaussian process modelling.

We study the case in which the physical system is the function $x \to x^2$ on $[0,1]$. The computer model is
$f_{mod}(x,\bbeta) = \beta_0 + \beta_1 x$. We assume that the covariance function of the model error is known and has the Gaussian form
$C_{mod}(x - y) = \sigma^2 \exp{ \left( - \frac{|x-y|^2}{l_c^2} \right) }$ with $\sigma = 0.3$ and $l_c = 0.5$.
There are three observations, noiseless, for experimental points $0.2$, $0.5$ and $0.8$. 

\begin{figure}[!h]
\begin{center}
\begin{tabular}{cc}
\includegraphics[angle=0,width=6cm,height=4cm]{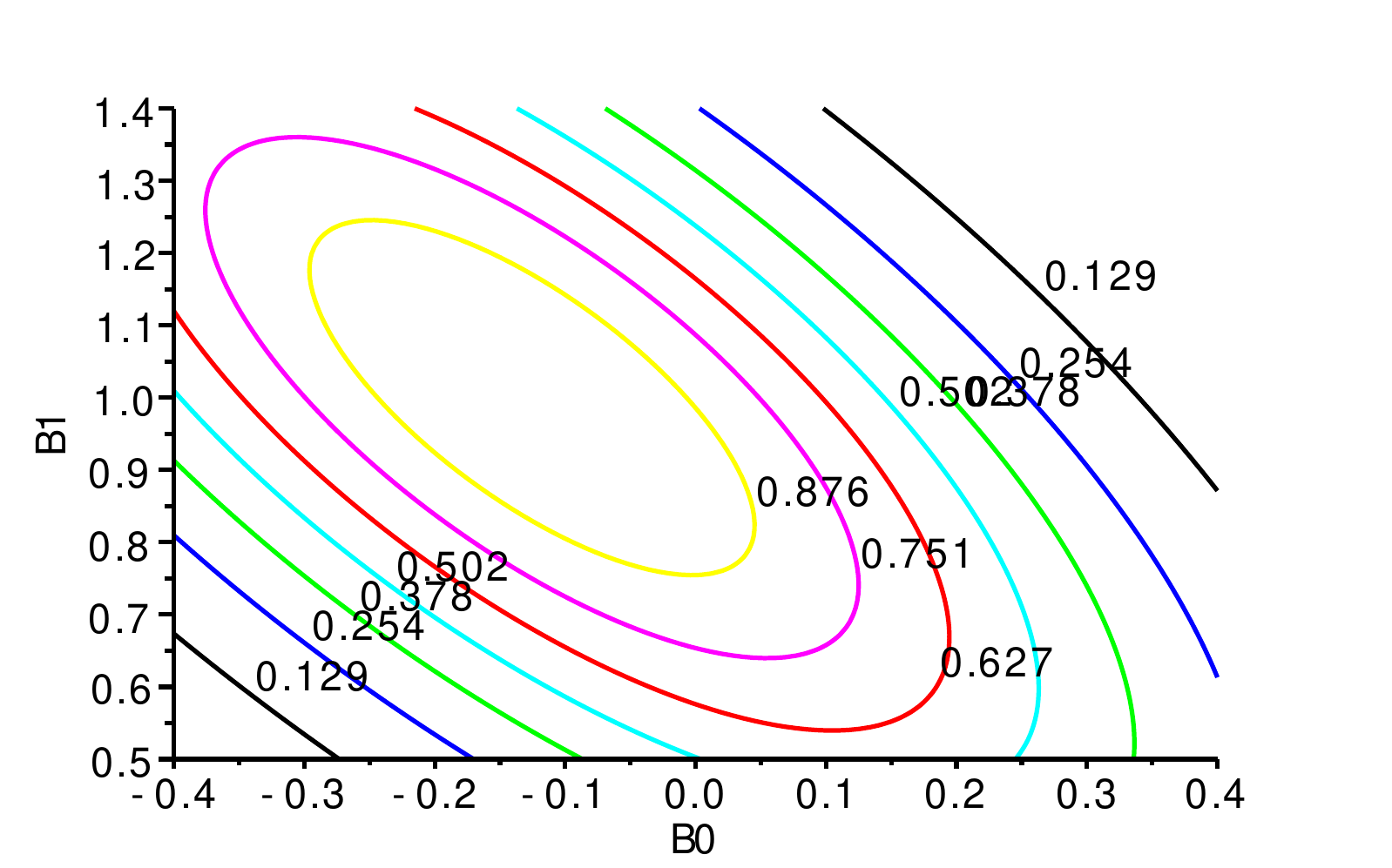}
&
\includegraphics[angle=0,width=5cm,height=4cm]{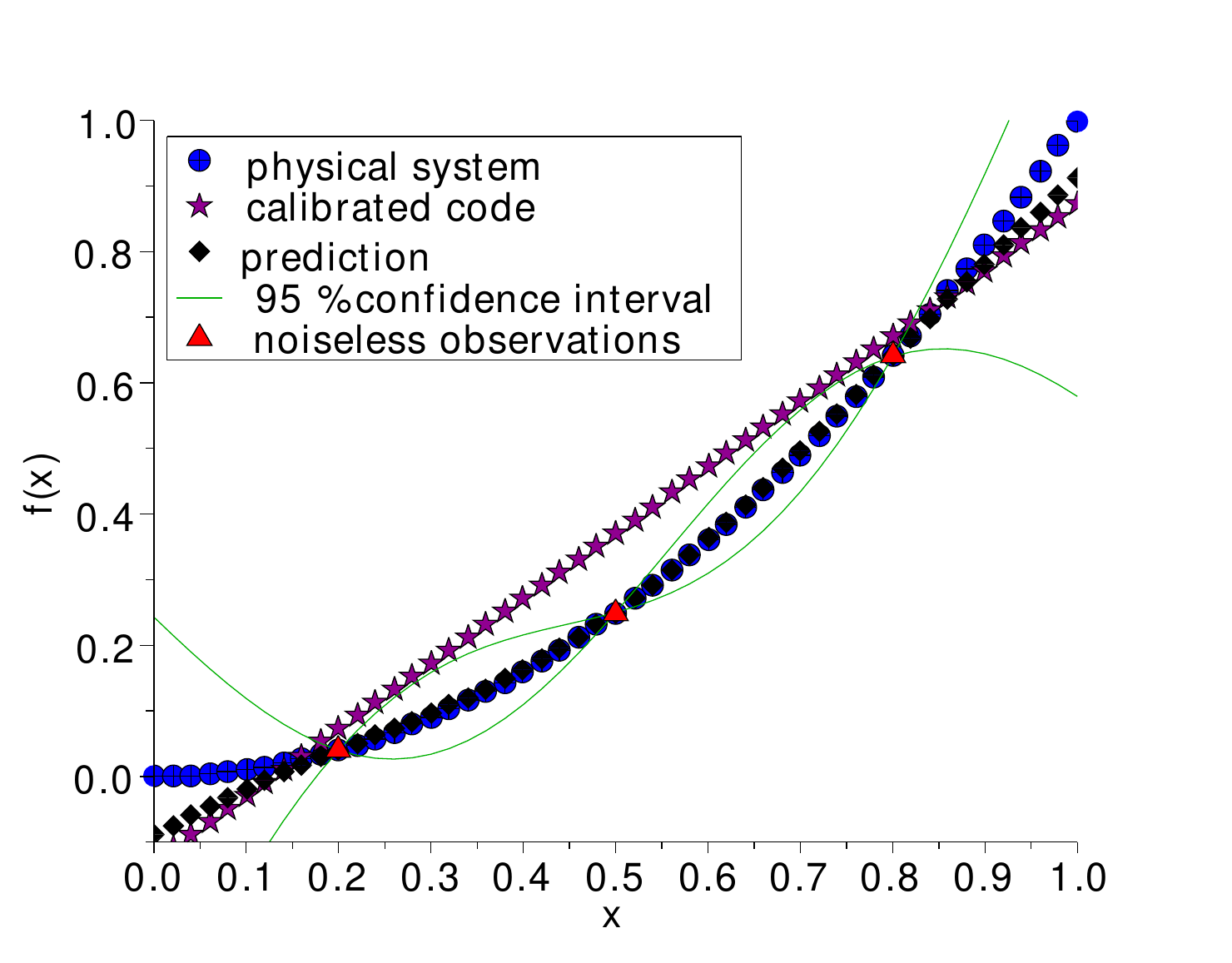}  \\
\end{tabular}
\end{center}
\caption[Illustration of calibration and prediction in the no prior information case]{ \doublespacing Manuscript title: Calibration and improved prediction of computer models by universal Kriging. 
Authors: Fran\c cois Bachoc, Guillaume Bois, Josselin Garnier and Jean-Marc Martinez. \\
\\
 Calibration and prediction in the no prior information case. Left: Iso-density curves of the probability density function for the estimation of $\bbeta$, given by \eqref{eq: calSIA}
and \eqref{eq: calSIA2}.
Right: Calibrated line \eqref{eq: calSIA}, real parabola, prediction \eqref{eq: predSIA} and $95\%$
confidence intervals of the form $[\hat{y}(\bx_{new}) - 1.96 \hat{\sigma}(\bx_{new}),\hat{y}(\bx_{new}) + 1.96 \hat{\sigma}(\bx_{new}) ]$.}
\label{fig: cal_freq}
\end{figure}

The results in the no prior information case are presented in figure \ref{fig: cal_freq}. We first see that there is a negative correlation in the estimation
of $\bbeta$. This correlation can be interpreted. Indeed if $\beta_0$, the value at $0$ of the line $ x \to \beta_0 + \beta_1 x$ is increased, then,
for the line to remain close to the parabola $x \to x^2$, the slope of the line ($\beta_1$) must be decreased. Furthermore an important remark is that
the calibrated line is above and does not go through the three observation points. This is surprising at first sight, all the more since the least square estimator of
section \ref{section: introduction} would go through the three points. This is because, as it is shown in \eqref{eq: predSIA}, the calibrated line is not intended to constitute a predictive model
of the parabola. Indeed it is completed by the inferred model error from the three observation points.  We see in figure \ref{fig: cal_freq}
that the prediction curve approximates almost perfectly the parabola. Let us also notice that in the extrapolation region ( {$0 \leq x \leq 0.2$ and $ 0.8 \leq x \leq 1 $} ),
the calibrated line approximates better the parabola than a line which would go between the three observation points. 

Hence, the inference of the model error improves
the prediction capability of the calibrated computer model. This is all the more true as the physical system is predicted closer to the experiments. In extrapolation,
the model error cannot be precisely inferred from the available observations and the inferred model error in \eqref{eq: predSIA} is hence very close to zero.
Hence, in extrapolation, the prediction is made using the calibrated computer model only. 
This is as expected, because when one cannot statistically improve the prediction of the computer model, a conservative choice is to rely only to physical knowledge.
Finally, we see that the confidence intervals (whose lengths are four times the standard deviations \eqref{eq: predSIA2} which corresponds to $95 \%$ confidence)
have length zero at the points where the noiseless observations are done,
and that this length increases when one moves away from
observation points. This shape of the confidence intervals is classical in Kriging with noiseless observations. 

We also consider the prior information case with
\[
\bbeta_{prior} = \left(  \begin{array}{c}
0.2   \\
1   \end{array}
\right),
\;\;
\Q_{prior} = \left(  \begin{array}{cc}
0.09 & 0   \\
0 & 0.09   \end{array}
\right).
\]

\begin{figure}[!h]
\begin{center}
\begin{tabular}{cc}
\includegraphics[angle=0,width=6cm,height=4cm]{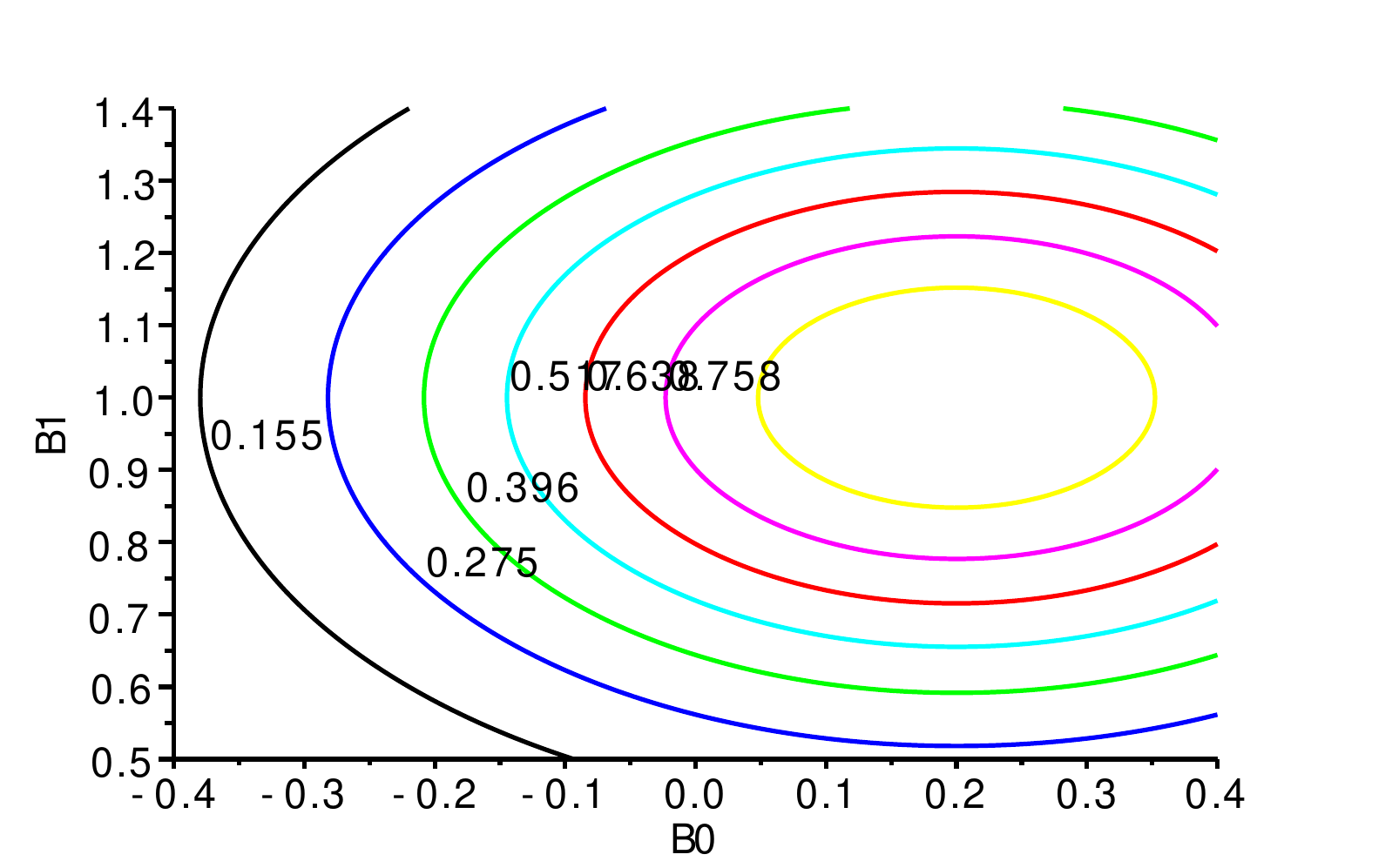}
&
\includegraphics[angle=0,width=6cm,height=4cm]{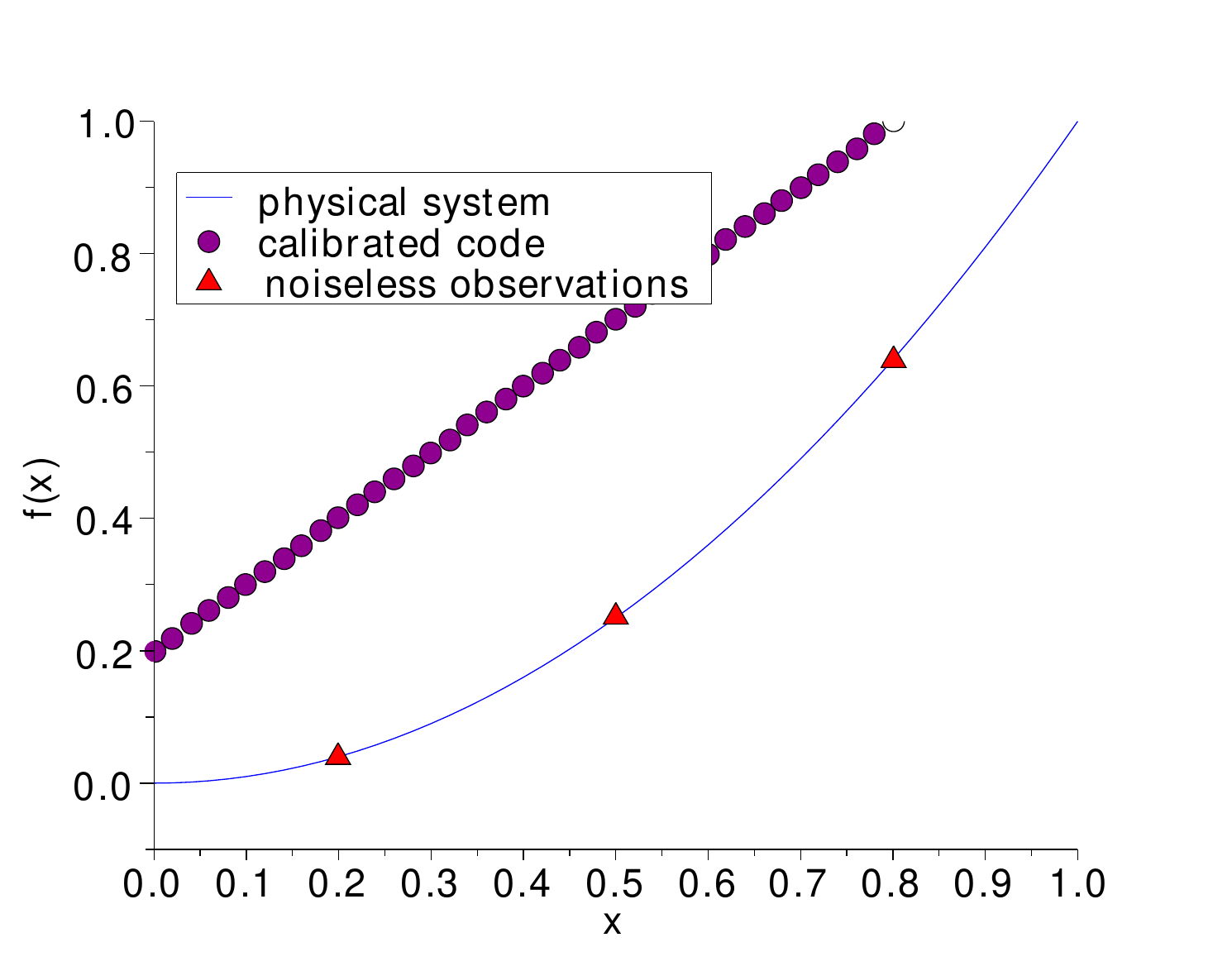}  \\
\includegraphics[angle=0,width=6cm,height=4cm]{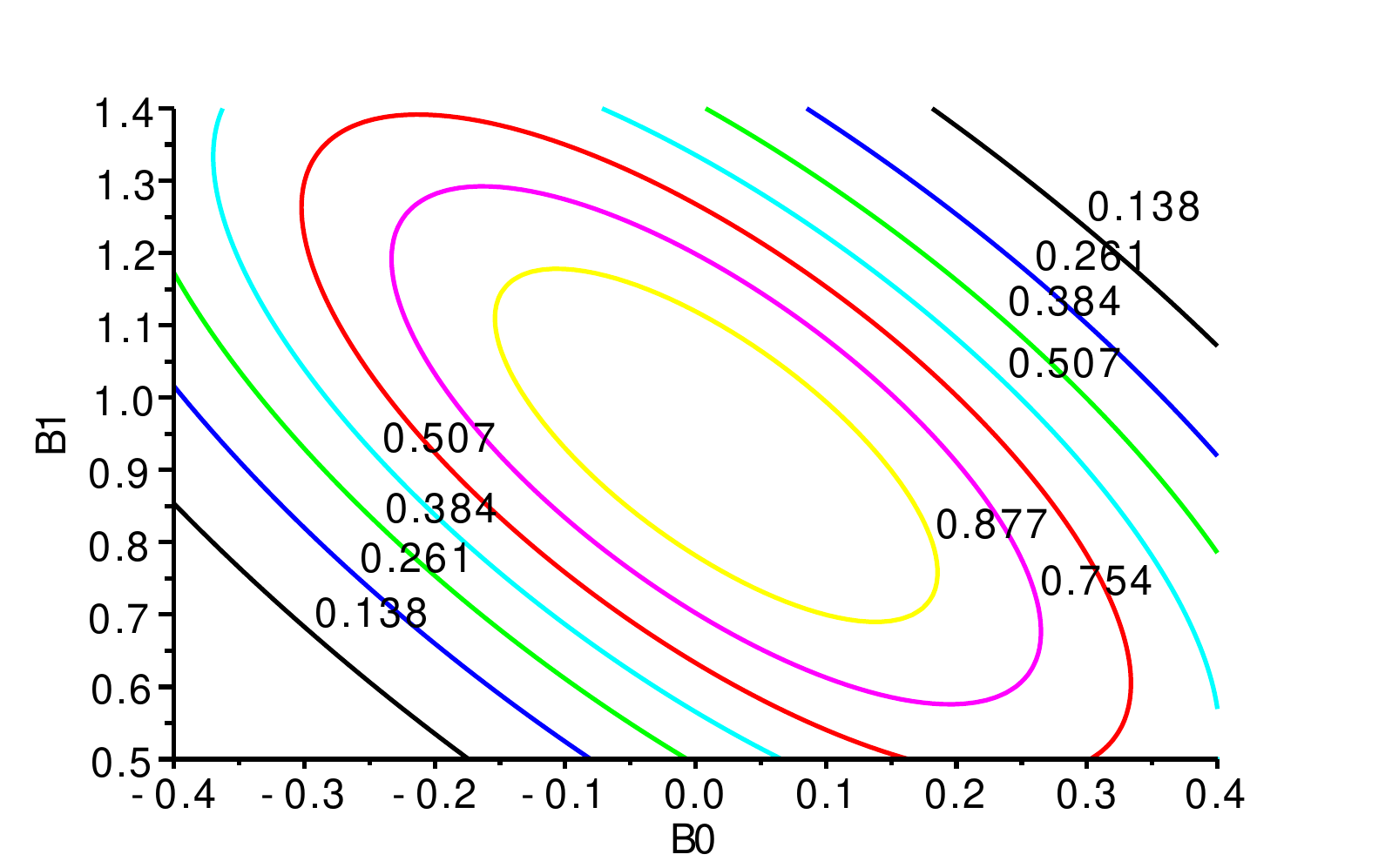}
&
\includegraphics[angle=0,width=6cm,height=4cm]{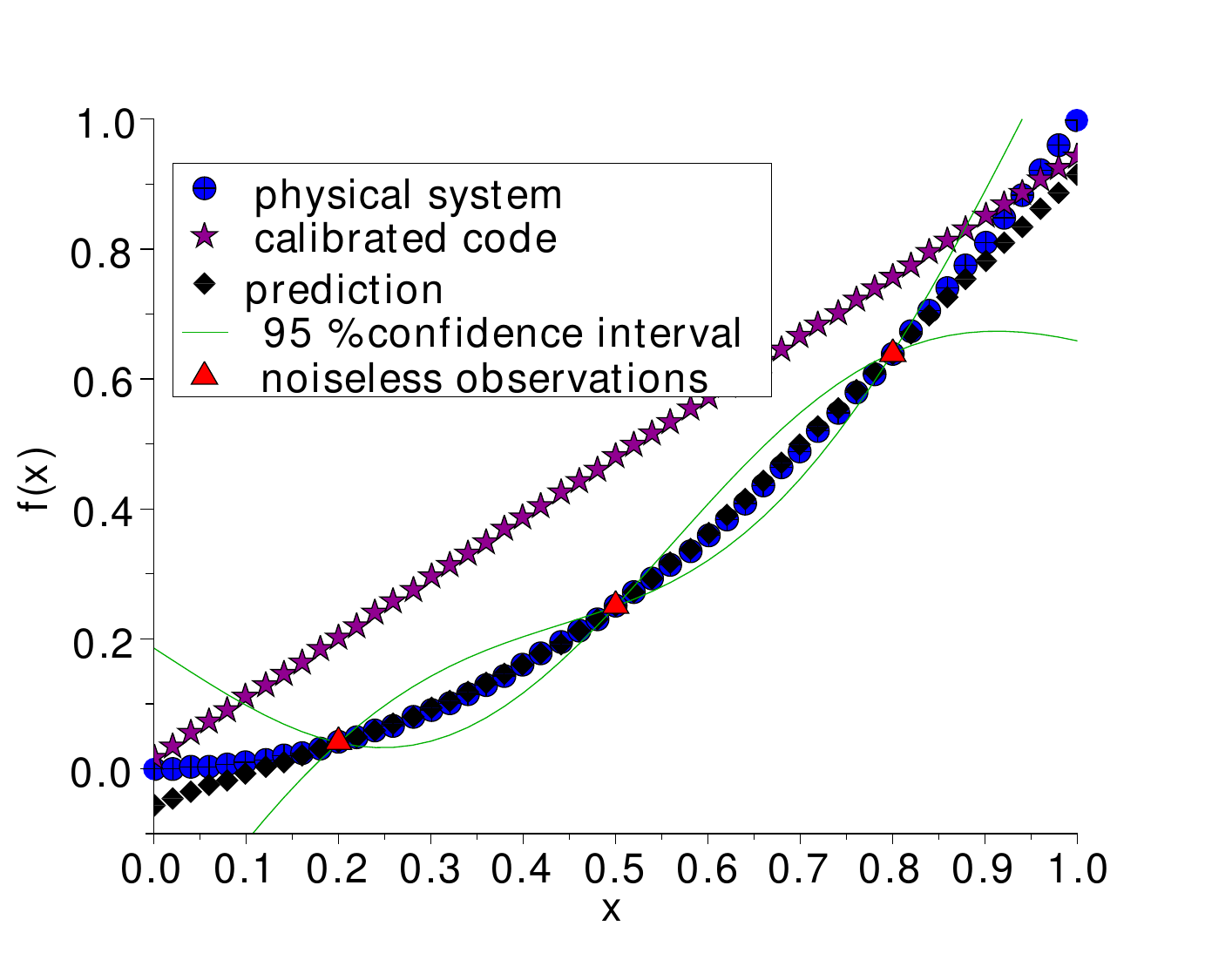}
\end{tabular}
\end{center}
\caption[Illustration of calibration and prediction in the prior information case]{ \doublespacing Manuscript title: Calibration and improved prediction of computer models by universal Kriging. 
Authors: Fran\c cois Bachoc, Guillaume Bois, Josselin Garnier and Jean-Marc Martinez. \\
\\
Calibration and prediction in the prior information case. Top left: iso-density curves of the prior probability density function of $\bbeta$.
Bottom left: iso-density curves of the posterior
probability density function of $\bbeta$ given by \eqref{eq: calAIA} and \eqref{eq: calAIA2}.
Right: Calibrated line with the prior (top) and posterior (bottom, \eqref{eq: calAIA}) mean values for $\bbeta$,
real parabola, prediction \eqref{eq: predAIA} and $95\%$ confidence intervals \eqref{eq: predAIA2}
of the form $[\hat{y}(\bx_{new}) - 1.96 \hat{\sigma}(\bx_{new}),\hat{y}(\bx_{new}) + 1.96 \hat{\sigma}(\bx_{new}) ]$.}
\label{fig: cal_bay}
\end{figure}

The results for this case are shown in figure \ref{fig: cal_bay}. Looking on the right plot, we can see that, from the prior $\bbeta$ to the posterior $\bbeta$, the line goes
substantially closer
to the three observation points. Nevertheless, it is not as close as in the no prior information case. This is a classical case in the prior information
case (as well as in Bayesian statistics), when the observations and the prior judgment are in disagreement, the posterior $\bbeta$ is a compromise
between the observations and the prior judgment. Looking on the left plot, we see that a negative correlation between the two components of $\bbeta$ appears
in the posterior law of $\bbeta$.

Finally, the prediction of the physical system, and the confidence intervals are similar to the no prior information case.

To conclude on the illustration on analytical functions, we see that the Gaussian process modelling has the potential to both improve the prediction
capability of the computer model and correctly assess the resulting uncertainty. In section \ref{section: application}, we confirm this on the computer model
FLICA 4 \cite{FDPFCCANMNA}, a thermal-hydraulic code relevant to core thermal-hydraulic transient and steady state analysis.
Before this, we give general practical recommendations concerning the use of the Gaussian process modelling.

\subsection{General recommendations for the Gaussian process modelling} \label{subsection: recommendation}

The first important point is that, as stated in section \ref{section: introduction}, the method presented here
does not address the complex field of code verification.
As a consequence, discretization or numerical parameters,
like the length or volume of a node, shall not be considered as model parameters or treated by the present method.

Another important point is the linear approximation of \eqref{eq: linear_approximation}. If the main objective is to achieve a precise enough prediction of
the physical system, and not to calibrate the computer model, then it is not a problem if the computer model is not linear with respect to its model
parameters.
Indeed, the linear approximation boils down to modelling the Gaussian process $Z$ in
\eqref{eq: mod_proc} as the model error of the linearized computer model in \eqref{eq: linear_approximation}.
In the prediction formulas \eqref{eq: predSIA} and \eqref{eq: predAIA}, we see that the statistical correction can compensate for
the linear approximation error of the code. This fact is confirmed in section \ref{section: application}
for the thermal-hydraulic code FLICA 4.
The linear approximation yields a much cheaper method than similar non-linear methods \cite{FVCM,CFDCSCP}, that may need to use
Markov Chain Monte Carlo (MCMC) methods, and possibly to approximate the computer model by a surrogate model in both the $\bx$ and $\bbeta$ domain.
Now, if calibration in itself is one of the main objectives,
one should act with caution with respect to the linear approximation. In this case, we advise to run a sensitivity analysis first to check the
linearity assumption (e.g the Morris method \cite{FSPPCE}). If the linearity assumption is infirmed,
then we recommend to proceed in two steps. First, a non-linear calibration should be carried-out, like the least square calibration or
a Bayesian calibration \cite{CFDCSCP}. Then, the model parameters should be fixed to their calibrated values, or a very narrow prior, centered around
these values, should be used, before using the present method.

Concerning the computation of the derivatives with respect to the model parameters $\bbeta$, two cases are possible. First the code can already
provide them, by means of the Adjoint Sensitivity Method \cite{SUAT}.
Similarly, automatic differentiation methods can be used on the source file of the code and
yield to a differentiated code \cite{Hascoet2004}. If these kinds of methods are not available, finite differences are necessary to
approximate the derivatives. Our main advice here is not to use a too small variation step.
Indeed, on the one hand, if the code is approximatively linear with respect to the model parameters, a too large variation step will
provide a good estimate of the derivatives anyway, whereas a too small variation step can yield numerical errors.
On the other hand, if the code is not approximatively linear, the linear approximation should not be used
for calibration. For prediction, the model error compensates for the linear approximation error as well as for the error in calculating the derivatives.

The fourth important point is that extrapolation is not recommended. This is a general advice for all Kriging models. The experimental results
should be made in the prediction domain of interest. Hence, for example, Kriging methods are not advisable to address scaling issues, that intrinsically ask
to extrapolate experimental results from one scale to another.

When dealing with more complex systems than the one of section \ref{section: application}, such as system-thermal hydraulics,
one may deal with high-dimensional problems, either with respect
to the number of experimental conditions (dimension of $\bx$) or to the number of model parameters (dimension of $\bbeta$).
The dimension of $\bx$ is a potential problem. A common rule of thumb for Kriging models is that one should have $n \geq 10 dim(\bx)$.
Note that screening methods exist and allow to select only the most impacting experimental conditions \cite{EMMCCCGP}. If the number of experiments
is really too small compared to the number of experimental conditions, our opinion is that it is not possible to take into account the model error correctly,
so that only the calibration should be carried-out.
If $\bbeta$ is high-dimensional, we advise, either to use a full Bayesian framework as described in section \ref{subsection: RMLE}, or
to select only the most important model parameters (from physical expertise), and to fix the other
model parameters at their nominal values. For example, in section \ref{section: application}, the less important parameters $a_l$, $C_f$, $n$ $d$
are fixed to their nominal values. In this case the process modeling the model error also compensates for the error made by freezing these parameters.

\section{Application to the thermal-hydraulic code FLICA 4} \label{section: application}

\subsection{Presentation of the thermal-hydraulic code FLICA 4 and of the experimental results}

The experiment consists in measuring the pressure drop in an ascending pressurized flow of liquid water through a tube that can be electrically heated.
This paper focuses on the frictional pressure drop ($\Delta P_{fric}$) in a single phase flow.

\paragraph{The thermal-hydraulic code FLICA 4}

The mathematical model for $\Delta P_{fric}$ is given by the local equation
\begin{equation} \label{eq: deltaPfro}
 \Delta P_{fric} = \frac{H}{2 \rho D_h} G^2  f_{iso} f_h.
\end{equation}
In \eqref{eq: deltaPfro}, each quantity is local. \eqref{eq: deltaPfro} is hence numerically integrated in space and time by the thermal-hydraulic code FLICA 4.
In \eqref{eq: deltaPfro}, $H$ is the friction height, $\rho$ is the density, $D_h$ is the hydraulic diameter, and  $G$ is the flowrate. 
$f_{iso}$ and $f_h$ are the friction coefficients respectively in the isothermal and heated flow regimes. The isothermal regime is defined by the
temperature of the liquid being uniformly equal to the wall temperature. On the other hand, the heated flow regime is characterized by a heat flux imposed on the test section and thus a varying liquid temperature. In this work, we focus on the single phase case, and we study the isothermal and heated flow subcases. 

The friction coefficient in the isothermal regime is 
\begin{equation} \label{eq: fiso}
f_{iso} = 
\begin{cases}
 \frac{a_l}{Re} & \text{if $Re <Re_l$}  \\
  \frac{a_t}{Re^{b_t}} & \text{if $Re_t < Re$}  \\
 \frac{a_l}{Re} \frac{Re_t - Re}{Re_t - Re_l} + \frac{a_t}{Re^{b_t}} \frac{Re - Re_l}{Re_t - Re_l} & \text{if $Re_l < Re < Re_t$}  \\
\end{cases}
\end{equation}
where $R_e = \frac{G D_h}{\mu}$ is the Reynolds number and $\mu$ is the viscosity. The limiting values $Re_l$ and $Re_t$ for the Reynolds number
are defined according to the literature and represent the limits of the transition regime between laminar and turbulent flows.
$a_l$, $a_t$ and $b_t$ are parts of the model parameters of the thermal-hydraulic code FLICA 4. They are the three components of the vector $\bbeta$
of model parameters in the isothermal regime. 

The friction coefficient in the heated flow regime is a correction factor expressed as
\begin{equation} \label{eq: fh}
 f_h = 1 - \frac{P_h}{P_w} \frac{C_f (T_w - T_b)}{1 + d \left( \frac{T_w + T_b}{2 T_0} \right)^n }
\end{equation}
where $P_h$ and $P_{w}$ are the heated and wetted perimeters, $T_w$ is the wall temperature, $T_b$ is the bulk temperature, and
$T_0 = 100^\circ C$
is a normalization temperature. $C_f$, $n$ and $d$ are the three components of the vector $\bbeta$ of model parameters in the heated flow case.
Finally, note that tests with no heat flux (isothermal tests) result in $T_w = T_b$, therefore the correction factor $f_h$ is equal to $1$, as expected.

\paragraph{The experimental results}

Several experimental tests have been conducted in order to calibrate FLICA 4 friction model. These tests have been used in previous calibration studies. The database
is composed of $n_i$ measurements under isothermal conditions, and $n_h$ measurements for heated tests. An experimental condition $\bx$ consists
in geometrical data (the channel width $e$, the hydraulic diameter $D_h$, and the friction height $H_f$) and in thermal-hydraulic conditions (the outlet
pressure $P_o$, the flowrate $G_i$, the wall heat flux $\phi_w$, the inlet liquid enthalpy
$h_i^l$, the thermodynamic title $X^i_{th}$, and the inlet temperature $T_i$). For each test the pressure drop due to friction $\Delta P_{fric}$ is
measured.

\subsection{Settings for the study}

\paragraph{Objectives}

We carry out the Gaussian process modelling method on the thermal-hydraulic code FLICA 4 in the isothermal and heated flow regimes.
We limit the calibration part of the study to
the parameters $a_t$ and $b_t$. That is to say, we enforce the parameter $a_l$ of the isothermal model, and the parameters $C_f$, $n$ and $d$ of the
heat correction model to their nominal values, computed in previous calibration studies. Indeed, the parameters $a_t$ and $b_t$
are the most influent parameters for the thermal-hydraulic code FLICA 4.

We work in the prior information case (calibration given by \eqref{eq: calAIA}). From previous
calibration studies, we have $\bbeta_{prior} = ( 0.22 , 0.21)^t$.
$\Q_{prior}$ corresponds to a $50\%$ uncertainty and is chosen diagonal with diagonal vector $(0.11^2,0.105^2)^t$.
Hence, this prior is rather large,
so that the calibration essentially depends on the experimental results.

An important point is that the two categories of experimental conditions (control and environment variables, see section \ref{section: introduction}) are
not equally represented in the experimental results.  The category of the control variables consists of the channel width $e$, the hydraulic diameter $D_h$, and the friction
height $H_f$. The category of the environment variables consists of the outlet pressure $P_o$, the flowrate $G_i$, the wall heat flux $\phi_w$, the liquid enthalpy
$h_i^l$, the thermodynamic title $X^i_{th}$, and the inlet temperature $T_i$. The $n_i+n_h$ experiments are divided into eight campaigns. Within a campaign, the control variables
remain constant, while the environment variables are varying. Hence, we only dispose of eight different control variables triplet. This means that, from the point of view of the
prediction given by the Gaussian process model \eqref{eq: predAIA}, it is a very unlikely that the prediction of the calibrated code is significantly improved when
considering new control variables. We experienced that, when predicting for new
control variables, the Gaussian
process method does not damage the predictions given by the nominal calibration of the thermal-hydraulic code FLICA 4 but it does not significantly improve it. However,
as we see next, we can give significantly improved predictions for observed control variables and new environmental variables.

To conclude, this study follows the double objective of calibration and prediction, in the prior information case for the parameters $a_t$ and $b_t$. Concerning the prediction,
the objective is to predict for experienced control variables and new environment variables.  

\paragraph{On the different covariance functions}

The environment and control variables listed above are not independent. Hence, it would be redundant to incorporate all of them in the covariance
function. One possible minimal set of environment and control variables is the set $(G_i,P_o,h_i^l,\phi_{\omega},H_f,D_h)$.
For this set, we will use the covariance function $C$, with $C$ being one of the four covariance functions of page \pageref{itemize: covarianceFunctions}.

To summarize, we represent the experimental conditions of an experiment by $\bx= (G_i,\phi_w,
h_i^l, P_o,H_f,D_h)$. The covariance function is
$C_{mod}(\bx^{(1)},\bx^{(2)}) = \sigma^2 C\left( \bx^{(1)} , \bx^{(2)} \right)$,
with $C$ being either, the exponential, the Mat\'ern $\frac{3}{2}$, the Mat\'ern $\frac{5}{2}$ or the Gaussian correlation function of
section \ref{itemize: covarianceFunctions}.
The hyper-parameters to be estimated are the variance $\sigma^2$ and the six correlation lengths $l_{c,1},...,l_{c,6}$.

Finally, we consider that the covariance matrix of the measurement error process is $\R_{mes} = \sigma_{mes}^2 \I_n$, with $\sigma_{mes} = 150 Pa$
provided by the experimentalists.

\paragraph{Cross Validation}

It is well known, in the general framework of statistical prediction, that the quality of a predictor should not be evaluated on the data that
helped to build it \cite{ESLDMIP} chapter 7. This is particularly true for the Gaussian process model, since it is based on the Kriging equations, that yields
an interpolation of the observations when there is no measurement error. When a rather limited number of observations is available, as is the case here, Cross Validation
is a very natural method to assess the predictive capability of a prediction model. In our case, we are interested in the two following quality criteria
for the Gaussian process predictor,
\begin{equation} \label{eq: RMSE_criteria_approx}
 RMSE^2 = \frac{1}{ n} \sum_{i_c = 1}^{n_c} \sum_{\bx \in C_{i_c} }  ( \hat{y}_{ \overline{C}_{i_c} }(\bx) - y_{obs}(\bx) )^2 
\end{equation}
and
\begin{equation} \label{eq: IC_criteria_approx}
 IC = \frac{ 1}{ n} \sum_{i_c = 1}^{n_c} \sum_{\bx \in C_{i_c} }
\mathbf{1}_{ | \hat{y}_{ \overline{C}_{i_c} }(\bx) - y_{obs}(\bx) | \leq 1.64 \left( \hat{\sigma}(\bx) \right)_{ \overline{C}_{i_c} } }. 
\end{equation}
In \eqref{eq: RMSE_criteria_approx} and \eqref{eq: IC_criteria_approx}, we use a $K$-fold Cross Validation procedure, with $K=10$. To do this, we partition the set of $n$ experiments into
$n_c = 10$ subsets $C_1,...,C_{n_c}$, each subset being well distributed in the experimental domain.
In \eqref{eq: RMSE_criteria_approx} and \eqref{eq: IC_criteria_approx}, $\overline{C}_{i_c}$ is the set of experimental conditions and observations that is the union
of the subsets
$C_1,...,C_{i_c-1},C_{i_c+1},...,C_{n_c}$. $\hat{y}_{ \overline{C}_{i_c} }(\bx)$ and $\left( \hat{\sigma}(\bx) \right)_{ \overline{C}_{i_c} }$
are the posterior mean and standard deviation of the predicted output at $\bx$ given the experimental data
in $\overline{C}_{i_c}$.
$[ \hat{y}_{ \overline{C}_{i_c} }(\bx) - 1.64 \left( \hat{\sigma}(\bx) \right)_{ \overline{C}_{i_c} } , \hat{y}_{ \overline{C}_{i_c} }(\bx) + 1.64 \left( \hat{\sigma}(\bx) \right)_{ \overline{C}_{i_c} } ]$
corresponds to a $90\%$ confidence interval.
It is
emphasized that at step $i_c$ of the Cross Validation, the Gaussian process model is built without using the experimental results of the class $C_{i_c}$.
Hence the important point is that, in the computation of the posterior mean and variance of the observed value at $\bx$,
this observed value is unused, for the estimation of the hyper-parameters as well as for the prediction formula. 

The Cross Validation presented here can yield a high computational cost, because one has to repeat the hyper-parameter estimation procedure $n_c$ times. When these
estimations are too costly, a simplified but approximate cross validation procedure is possible in which the hyper-parameters are estimated only once for all.
For this simplified version, Cross Validation is carried out only with respect to the prediction
formulas of \eqref{eq: predAIA} and \eqref{eq: predAIA2}. Let us notice that, in this context, there exists formulas \cite{CVKUN}
that allow to calculate the result of the Cross Validation procedure without actually calculating $K$ times the prediction formulas
\eqref{eq: predSIA}, \eqref{eq: predSIA2}, \eqref{eq: predAIA} and \eqref{eq: predAIA2}. These "virtual" Cross Validation formulas reduce even
more the Cross Validation computational cost.
Nevertheless, in our case, we are able to estimate the hyper-parameters at each step of the Cross Validation. Indeed, we have
a rather limited number $n$ of experimental results (the computation of the Restricted Likelihood is $O(n^3)$).

\subsection{Results}

\paragraph{Results in the isothermal regime}

In a first step, we consider the results in the isothermal and turbulent flow regime only. That is to say, the regime when $f_h = 1$ in \eqref{eq: deltaPfro}, and when
$Re > Re_t$ in \eqref{eq: fiso}.
We have $n_{it}$ experimental results.

The isothermal regime is characterized by no wall heat flux, $\phi_w = 0$. Hence, it is useless to include it in the covariance function, because it is
uniformly zero for all the experimental conditions. So, we only have five
correlation lengths out of six to estimate, which are $l_{c,1}$, $l_{c,3}$, $l_{c,4}$, $l_{c,5}$ and $l_{c,6}$ corresponding to $G_i$, $h_i^l$,
$P_o$, $D_h$ and $H_f$. 

\begin{figure}[!h]
\begin{center}
\includegraphics[angle=0,width=12cm]{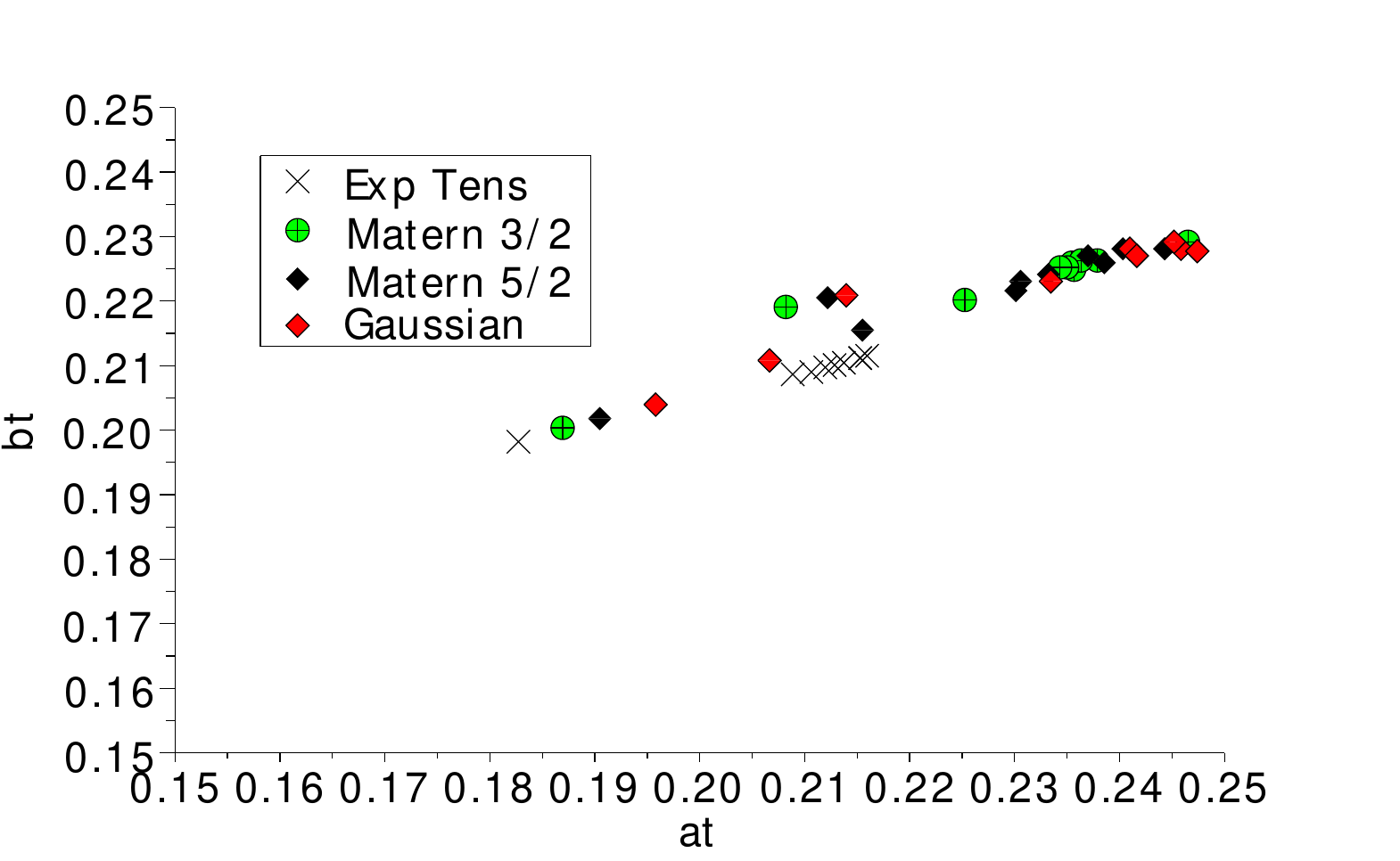}
\end{center}
\caption[Calibration in the isothermal regime]{ \doublespacing Manuscript title: Calibration and improved prediction of computer models by universal Kriging. 
Authors: Fran\c cois Bachoc, Guillaume Bois, Josselin Garnier and Jean-Marc Martinez. \\
\\
 Calibration in the isothermal regime. $10$-fold Cross Validation. Plot of the $n_c = 10$ posterior means \eqref{eq: calAIA} of $a_t$ and $b_t$ for the exponential, Mat\'ern $3/2$, Mat\'ern $5/2$ and
Gaussian covariance functions of section \ref{itemize: covarianceFunctions}.} \label{fig: IsoTurCal}
\end{figure}

On figure \ref{fig: IsoTurCal}, we plot, for the $10$-fold Cross Validation, the $n_c = 10$ posterior mean values of $a_t$ and $b_t$ for the four covariance functions of
page \pageref{itemize: covarianceFunctions}. The conclusions are that the Gaussian process calibration does not change significantly the nominal values $a_t = 0.22$ and
$b_t = 0.21$. Furthermore we do not notice significant differences concerning the choice of the covariance function for the calibration. Finally, we can observe
a high correlation in the posterior means of $a_t$ and $b_t$. This is confirmed in the $n_c$ posterior covariance matrix, where the correlation
coefficient is larger than $0.95$.

Concerning the prediction, we first compute the $RMSE$ and $IC$ criteria for the four covariance functions. Results are presented in table \ref{table: IsoTurPred}.
The first comment is that the predictive variances of \eqref{eq: predAIA2} are reliable, because they yield rather precise $90 \%$ confidence intervals. This is also
observed for Kriging, e.g in \cite{GEUKUP}. The second comment is that there is no significant difference between the different covariance functions.
This may be due to the amplitude of the measurement error, which makes insignificant the problem of the regularity of the covariance function. It is shown in
\cite{ISDSTK} Section 3.7 that, in a particular asymptotic context, even a small measurement error can have a significant effect on prediction errors. 

\begin{figure}[!h]
\begin{center}
\begin{tabular}{cc}
\includegraphics[angle=0,width=6.5cm,height=4.5cm]{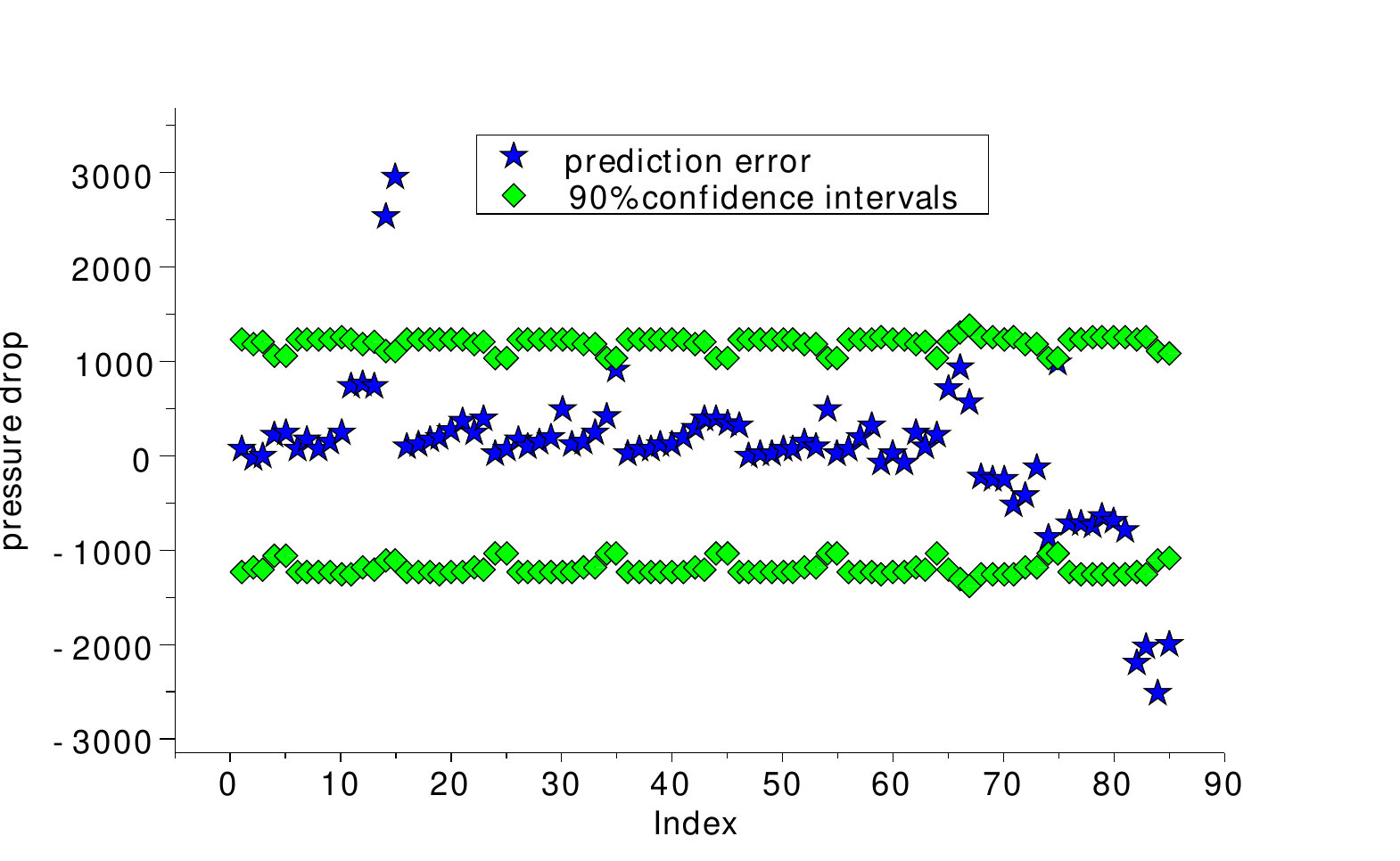}
&
\includegraphics[angle=0,width=6.5cm,height=4.5cm]{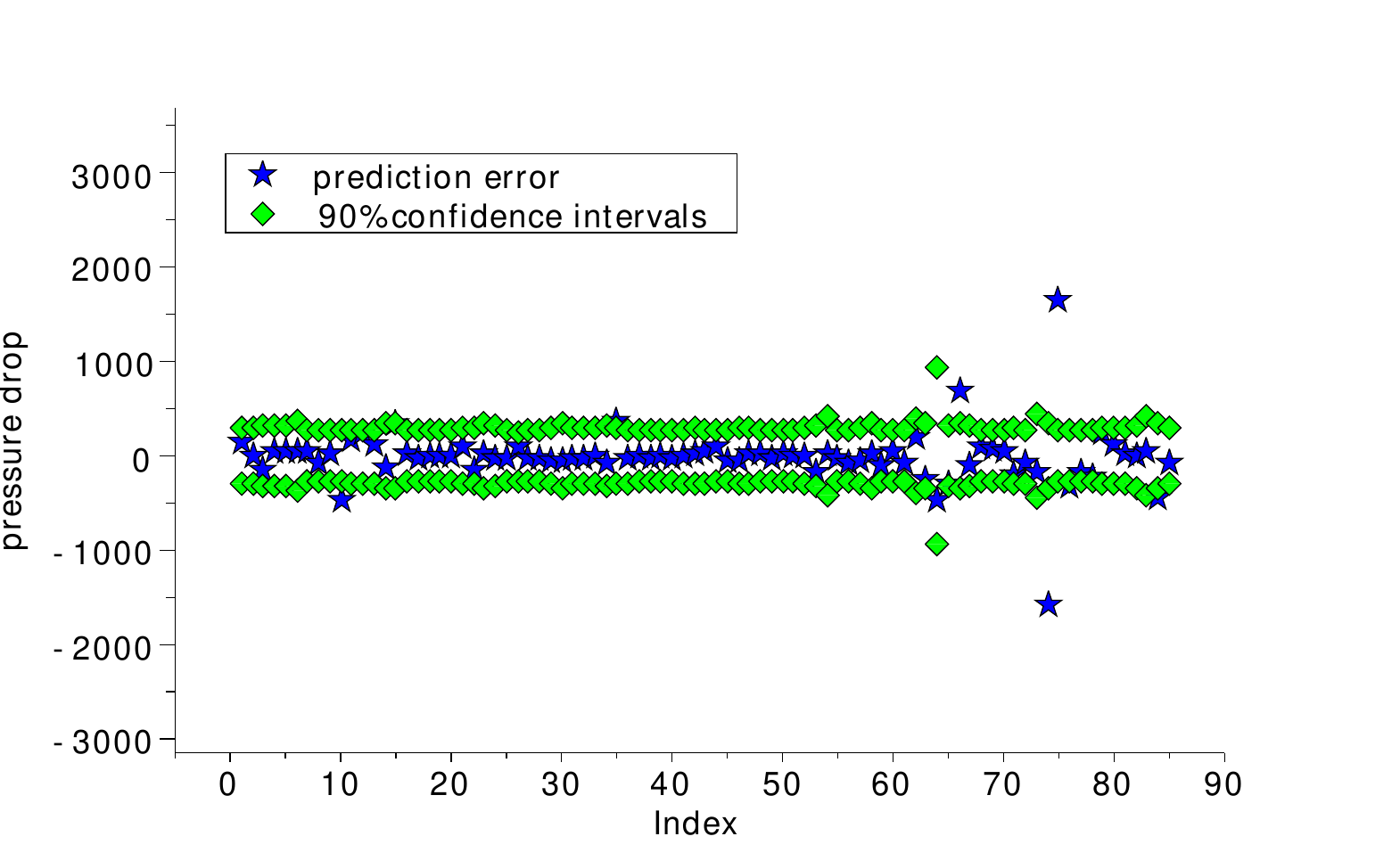}  \\
\end{tabular}
\end{center}
\caption[Prediction errors in the isothermal regime]{  \doublespacing Manuscript title: Calibration and improved prediction of computer models by universal Kriging. 
Authors: Fran\c cois Bachoc, Guillaume Bois, Josselin Garnier and Jean-Marc Martinez. \\
\\
 Prediction errors (observed values minus predicted values \eqref{eq: predAIA}) and $90 \%$ confidence intervals for these prediction errors,
derived by the calibrated thermal-hydraulic code FLICA 4 (left), and the Gaussian process
method (right). $90 \%$ confidence intervals are of the form $[ - 1.65 \hat{\sigma} ( \bx ) ,  1.65 \hat{\sigma} ( \bx ) ]$ with
$\hat{\sigma} ( \bx )$ given by \eqref{eq: predAIA2}. 
Plot with respect to the index of experiment.} \label{fig: IsoTurPred}
\end{figure}

We now present more detailed results for the Mat\'ern $\frac{3}{2}$ covariance function.
We first compare the Gaussian process predictions with the predictions given by the calibrated code alone. With the same Cross Validation procedure, the RMSE criterion for the
calibrated code alone is $RMSE =  741 Pa$. This is to be compared with a $RMSE$ around $300 Pa$ for the Gaussian process method.
Hence the inference of the model error process significantly improves the predictions of the code.
We illustrate this in figure \ref{fig: IsoTurPred}, where we plot, for each of the $n_{it}$ observations, the predicted values and confidence intervals with the
$10$-fold Cross
Validation method. The plots are done with respect to the experiment index. This index has physical meaning, because two experiments with successive indices are similar
(for instance, the experiences of a given campaign have successive indices). We first see that the Gaussian process modelling significantly reduces the prediction
errors, and that the confidence intervals are reliable. Then, we observe a regularity in the plot of the prediction error for the calibrated code, especially for
the largest indices. This regularity is not
present anymore in the error of the Gaussian process method. The conclusion is that the Gaussian process method detects a regularity in the error of the calibrated code, and
uses it to significantly improve its predictions.

Finally, in table \ref{table: IsoTurHyperPar}, we show the $n_c=10$ different estimations of $( \sigma^2 , l_{c,1} , l_{c,3}, l_{c,4} , l_{c,5} , l_{c,6} )$,
for the different steps of the Cross
Validation.
The first conclusion is the singularity at steps $5$ and $6$ of the Cross Validation. The
explanation is that, among the $n_{it}$ experimental results, there are two singular points that have very similar experimental conditions but
substantially different values for the quantity of
interest. These two points are in CV classes $5$ and $6$. Hence the estimation of the hyper-parameters in the CV steps
$1$, $2$, $3$, $4$, $7$, $8$, $9$, $10$, where this singularity is present in the data used for the estimation, is different
from the steps $5$ and $6$, where the singularity is absent.
On figure \ref{fig: IsoTurPred}, these two singular points yield the two largest prediction errors for the Gaussian process method.
Indeed, when one of them is in the test group,
the other is in the learning group. As the Gaussian process modelling principle is to assume a correlated model error, the quantity of interest of the
singular point
of the test group is (up to the measurement error) predicted by the quantity of interest of the singular point of the learning group.

The correlation lengths in table \ref{table: IsoTurHyperPar} correspond to normalized experimental conditions varying between $0$ and $1$.
Hence, the second conclusion is that the estimated
correlation lengths are rather large, corresponding to rather large scales of variations of the model error, as discussed for figure \ref{fig: IsoTurPred}.
When an estimated  correlation length is very large (larger than $10$), it is equivalent to assuming than the model error is independent of the
corresponding experimental condition.
The third
conclusion is that the estimations of the hyper-parameters can vary moderately among the Cross Validation steps. This is an argument in favor of reestimating the
hyper-parameters at each step of the Cross Validation, because this takes into account these variations.
Finally let us notice, that, for the Gaussian process model to be used for
new experimental conditions, the hyper-parameters are to be reestimated with all the observations.

\paragraph{Results in the single phase regime}

We now use all the experiments of the single phase regime (isothermal and heated flow regimes), that is to say $n=n_i+n_h$ experiments. Hence, we estimate six correlation
lengths for the six environment and control variables $G_i$, $\phi_w$, $h_i^l$, $P_o$, $D_h$ and $H_f$.

Concerning the prediction, we first compute the $RMSE$ and $IC$ criteria for the four covariance functions. Results are presented in table \ref{table: MonoPred}.
As in the isothermal case, we see that the predictive variances are reliable and that there is no significant difference between the four covariance functions. As for
the isothermal regime, we present in more details
the results for the the Mat\'ern $\frac{3}{2}$ covariance function.

\begin{figure}[!h]
\begin{center}
\begin{tabular}{cc}
\includegraphics[angle=0,width=6.5cm,height=4.5cm]{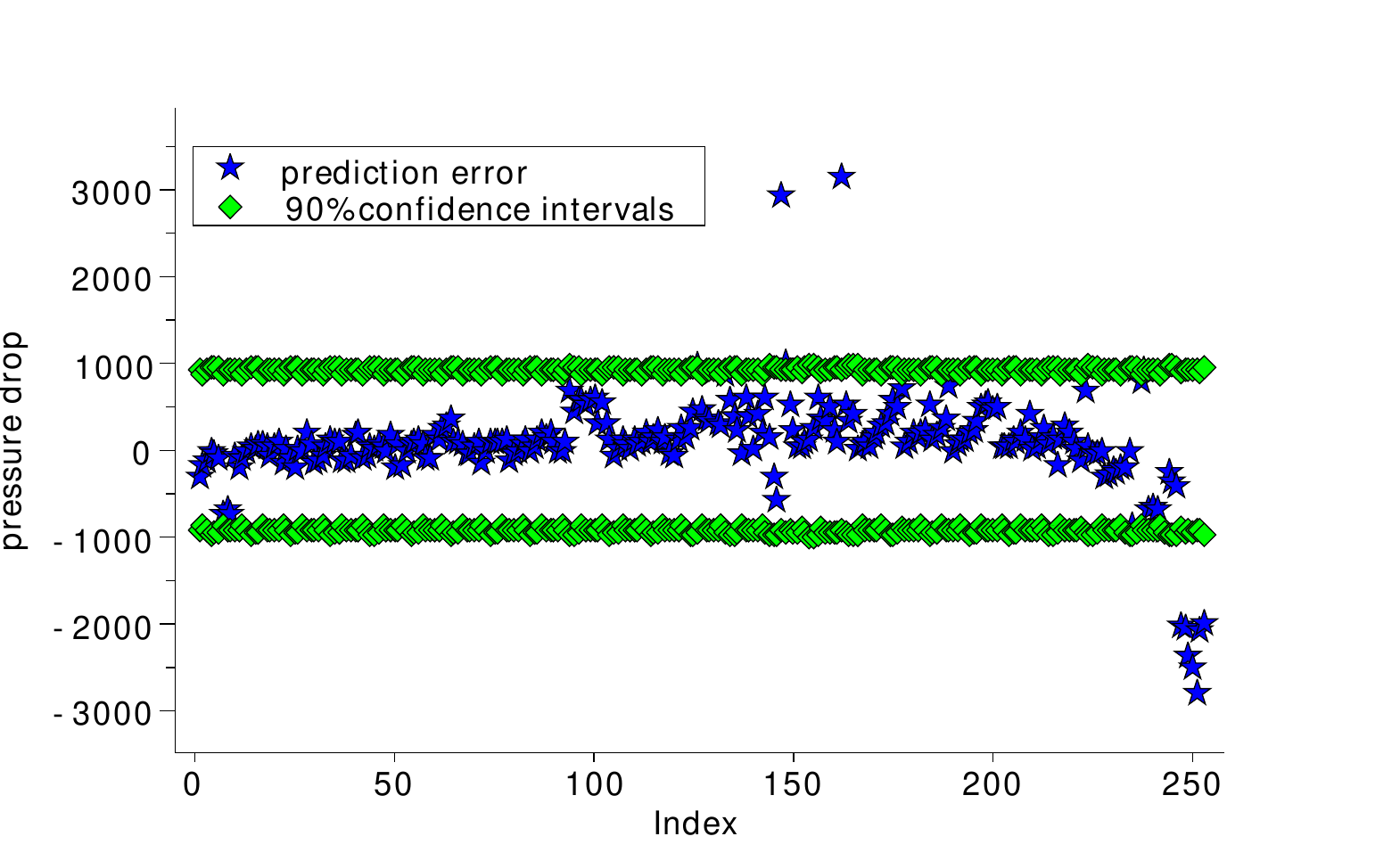}
&
\includegraphics[angle=0,width=6.5cm,height=4.5cm]{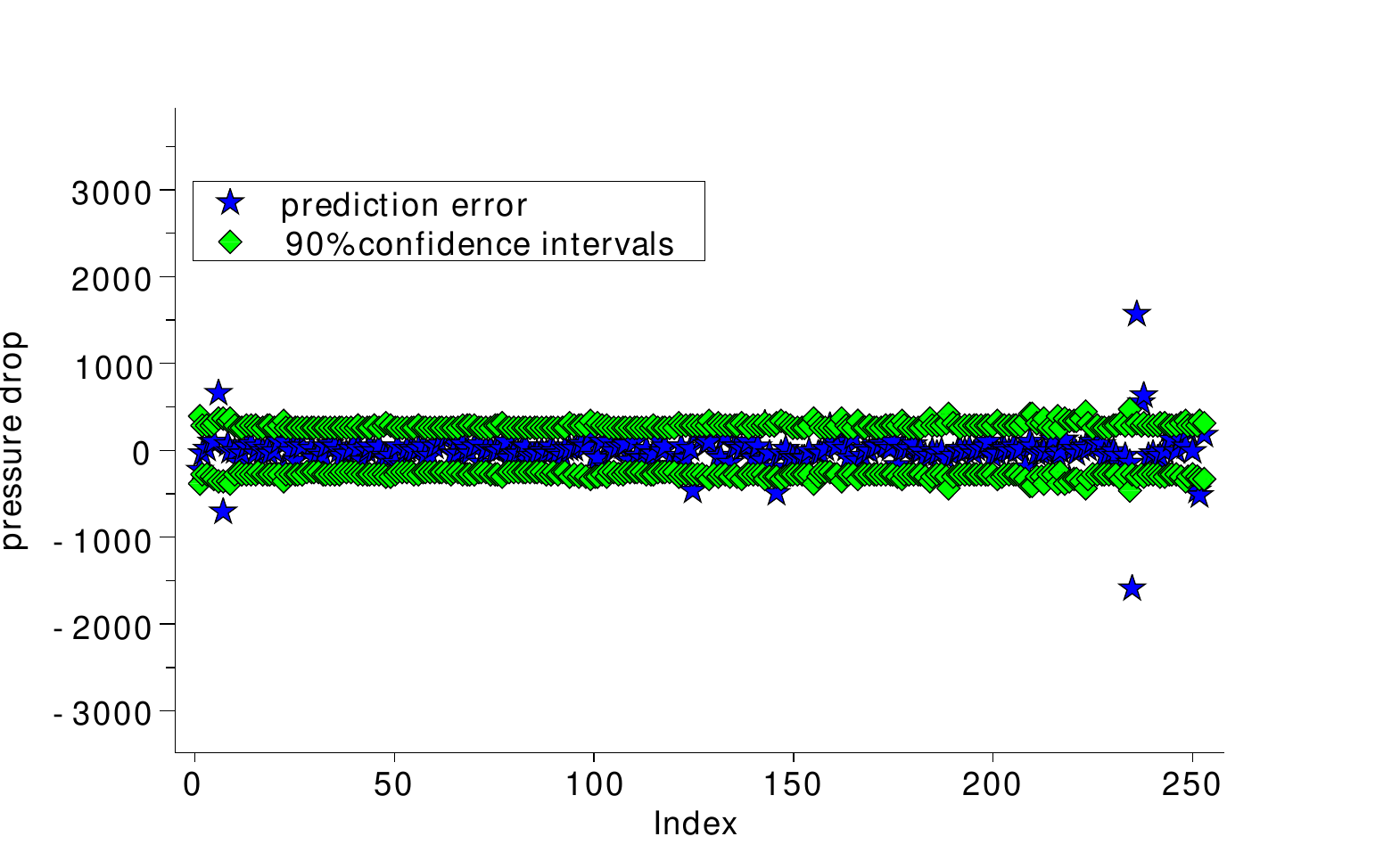}  \\
\end{tabular}
\end{center}
\caption[Prediction errors in the single phase regime]{  \doublespacing Manuscript title: Calibration and improved prediction of computer models by universal Kriging. 
Authors: Fran\c cois Bachoc, Guillaume Bois, Josselin Garnier and Jean-Marc Martinez. \\
\\
Same settings as in figure \ref{fig: IsoTurPred} but in the single phase regime.} \label{fig: MonoPred}
\end{figure}

With the same Cross Validation procedure, the RMSE criterion for the
calibrated code alone is $RMSE =  567 Pa$. This is to be compared with a $RMSE$ around $200 Pa$ of the Gaussian process method.
Hence the inference of the model error process significantly improves the predictions of the code, in the same way as in the isothermal regime.
We illustrate this in figure
\ref{fig: MonoPred}, where we plot the same quantities as in figure \ref{fig: IsoTurPred}. We obtain the same conclusion: the Gaussian process model
detects a regularity in the error of the calibrated code, and uses it to improve its predictions.

\paragraph{Influence of the linear approximation}

All the results above are obtained using the linear approximation of the thermal-hydraulic code FLICA 4 with respect to
$a_t$ and $b_t$. We have implemented the equivalent of the calibration and prediction formulas of
\eqref{eq: calAIA} and \eqref{eq: predAIA},
when the thermal-hydraulic code FLICA 4 is not considered linear with respect to
$a_t$ and $b_t$ \cite{BVCM}. Integrals in the $a_t,b_t$ domain were calculated on a $5 \times 5$ grid, which, to avoid bias, was also used
when the linear approximation of the thermal-hydraulic code FLICA 4 was used. Using the same $10$-folds CV procedure as before, in the
single phase regime, we obtain $RMSE = 197.8$ with the linear approximation and $RMSE = 196.9$ without the linear approximation (less than $1\%$ relative difference).
The posterior means of $a_t$ and $b_t$, along the different CV steps, have a Root Mean Square Difference of
$0.025$ (more than $10 \%$ relative difference), between the cases where the linear approximation was made or not.
Hence, this is an illustration of the general remark of section \ref{subsection: recommendation}: if the computer model is non-linear
with respect to its calibration parameters, it is the model error with respect to the linearized computer model that is inferred.
Thus, the predictions of the physical system are similar, whether or not the linear approximation is made.

\section{Conclusion}

In this work, a Gaussian process modelling method has been presented for computer model calibration and
improved prediction of the underlying physical system.
It is based on a modelling of the model error, which is the bias between the computer model and the physical system. 
A set of experimental results on the physical system is used, which enables to infer the model error for each new potential experimental point.
As a result, an improved prediction for the value of the physical system, and an associated confidence interval, are provided.

 The Gaussian process modelling method is carried out in two steps. In a first step, the covariance function of the model error is estimated, based
on the comparison between experimental results
and the computer model. In this paper, the estimation is done with the Restricted Maximum Likelihood method, although the
possibility of using other methods is discussed. This estimation step yields the main computational cost, since one needs to minimize a function involving a matrix inversion.
The size of this matrix is equal to the number of experimental results. 

Once the estimation is done, calibration and prediction can be carried out with closed form matrix-vector formulas. The calibration is the computation of the best
parameters for the computer model. Physical knowledge on the calibration parameters of the physical model can be taken into account in a Bayesian framework. The prediction
is the computation of a predicted value and of an associated confidence interval for each new potential experimental point. The predicted value is the sum of the
calibrated computer model and of a Gaussian inference of the model error. Hence, the calibrated computer model is completed by a statistical term. This statistical term is
based on the experimental results, and can significantly improve the predictions of the computer model. The closed form linear algebra formulas for calibration and
prediction rely on a linearization of the computer model, with respect to the model parameters, around a reference parameter.
These formulas can still be used when the linear approximation does not hold,
in which case, the calibration will be carried out on the linearized computer model, and the model error will incorporate the linear approximation error.
It is shown that the linear approximation has no consequence on the prediction, but shall be treated carefully if calibration is one of the main objectives.
The Gaussian process modelling of the model error can be carried out without linearization of the computer model
\cite{FVCM,CFDCSCP}, but this yields a much more costly computation.

The method is applied to the friction model of the thermal-hydraulic code FLICA 4, for which
the data of several experimental campaigns are available.
We evaluate the prediction capability of either the calibrated code alone or the Gaussian process modelling method. This evaluation is done rigorously using a ten-fold
Cross Validation on the experimental results.
It is shown that the error of the thermal-hydraulic code FLICA 4 can be divided by a factor between two and three. We also study different covariance functions
for the model error, and come
to the conclusion that, due to the measurement errors, the choice of the covariance function does not have significant influence on the prediction capability in this case.

Based on this case study, we believe the Gaussian process modelling of the model error to be promising in the field of computer model validation
for Nuclear Engineering, by its ability to complete a computer model with a statistical inference of the model error.

An interesting area of research is the implementation of this method to functional output computer models, arising for example
in the case of time dependent problem. In the general context of Kriging with functional output, two kind of methods exist.
The first solution is to consider a joint covariance structure, with respect to the inputs, and with respect to the functional output time or space parameter
\cite{Rougier2008}. The second solution is to use a low-dimensional representation of functional outputs, such as PCA or wavelets,
and to build a Kriging model for each of the coefficients of the representation.
The adaptation of these methods to the Gaussian process modelling of the model error, in a Nuclear Engineering context, may motivate further research.

\section{Acknowledgments}
We thank the two anonymous reviewers for their comments and suggestions, which
helped to improve the quality of the manuscript.

\FloatBarrier
\newpage

\begin{table}[!h]
\begin{center} 
\begin{tabular}{| c | c |  c |}
\hline
Covariance function &  $RMSE$ ($Pa$) & $IC$ \\ 
\hline
 exponential & $289.5$ & $0.93$ \\
\hline
 Mat\'ern $\frac{3}{2}$  & $296.2$ & $0.92$ \\
\hline
 Mat\'ern $\frac{5}{2}$ & $302.7$ & $0.89$ \\
\hline
 Gaussian & $310.8$ & $0.88$ \\
\hline
\end{tabular}
\end{center} 
\caption[Prediction results in the isothermal regime]{ \doublespacing Prediction results in the isothermal regime.\\
RMSE and IC criteria of \eqref{eq: RMSE_criteria_approx} and \eqref{eq: IC_criteria_approx}
obtained with a $10$-fold Cross Validation procedure, for the covariance functions presented in section \ref{itemize: covarianceFunctions}.
}
\label{table: IsoTurPred}
\end{table}

\FloatBarrier
\newpage

\begin{table}[!h]
\begin{center} 
\begin{tabular}{| c | c | c |  c | c | c |c|}
\hline
Cross Validation step & $\sigma$ &  $l_{c,1}$ & $l_{c,3}$  & $l_{c,4}$ &  $l_{c,5}$ & $l_{c,6}$ \\ 
\hline
  1 &  $2220$ &   $2.3$  & $4.0$ & $100$&              $0.40$ &  $53$  \\
  \hline
  2 &  $2100$  & $2.2$ &  $3.5$  &  $100$&              $0.40$  & $100$\\
   \hline
  3 & $2088$ &  $2.1$ &  $3.8$   &$100$&              $0.39$ &  $100$    \\   
   \hline
  4 & $2266$ &  $2.3$  &$2.0$ &   $100$&              $0.50$ &  $100$      \\      
   \hline
  5 & $4491$ & $3.4$&   $100$&              $24$ &   $1.36$ &   $100$        \\    
  \hline
  6 &  $1953$ &  $1.6$& $15$ &  $3.4$ & $7.7$ &  $0.6$\\
  \hline
  7 &  $2385$& $2.4$ &  $4.6$ &  $100$&              $0.44$ & $100$            \\
   \hline
  8 & $2436$ &$2.4$&  $4.8$ &$100$&              $0.45$  & $99$\\
   \hline
  9 & $2331$  & $2.4$ &$4.2$  &  $100$    &          $0.43$ &  $100$\\            
  \hline
  10 &  $2294$  &  $2.4$&  $3.8$ &  $100$&              $0.42$ & $100$\\
\hline
\end{tabular}
\end{center} 
\caption[Estimated hyper-parameters in the isothermal regime]{ \doublespacing Estimated hyper-parameters in the isothermal regime. \\
Estimated correlation lengths for the Mat\'ern $\frac{3}{2}$ covariance function of section \ref{itemize: covarianceFunctions},
for the $10$-fold Cross Validation procedure.
}
\label{table: IsoTurHyperPar}
\end{table}

\FloatBarrier
\newpage

\begin{table}[!h]
\begin{center} 
\begin{tabular}{| c | c |  c |}
\hline
Covariance function &  $RMSE$ ($Pa$) & $IC$ \\ 
\hline
 exponential & $202.2$ & $0.95$ \\
\hline
 Mat\'ern $\frac{3}{2}$  & $196.2$ & $0.95$ \\
\hline
 Mat\'ern $\frac{5}{2}$ & $ 196.9$ & $0.95$ \\
\hline
 Gaussian & $199.5$ & $0.94$ \\
\hline
\end{tabular}
\end{center} 
\caption[Prediction results in the single phase regime]{ \doublespacing Prediction results in the single phase regime. \\
Same setting as in table \ref{table: IsoTurPred}.}
\label{table: MonoPred}
\end{table}

\FloatBarrier
\newpage

\FloatBarrier
\newpage

\FloatBarrier
\newpage

\FloatBarrier
\newpage

\FloatBarrier
\newpage

\FloatBarrier
\newpage

\bibliographystyle{unsrt}
\bibliography{Biblio.bib} 

\end{document}